\documentclass[twocolumn]{aastex63}

\newcommand{\sm}{log(M_{\star}/M_{\odot})}
\newcommand{\neiii}{[Ne \textsc{iii}]$\lambda$3869}
\newcommand{\oii}{[O \textsc{ii}]$\lambda\lambda$3727,3729}
\newcommand{\oiii}{[O \textsc{iii}]$\lambda\lambda$4959,5007}
\newcommand{\auroral}{[O \textsc{iii}]$\lambda$4363}
\newcommand{\neon}{[Ne \textsc{iii}]}
\newcommand{\nii}{[N \textsc{ii}]$\lambda\lambda$6548,6583}
\newcommand{\ha}{H$\alpha$6563}
\newcommand{\hb}{H$\beta$4861}

\shorttitle{The MZR in Dwarf Galaxies}
\shortauthors{Pharo et al.}
\graphicspath{{./}{figures/}}

\begin{document}

\title{Dwarf Galaxies Show Little ISM Evolution from $z\sim1$ to $z\sim0$: a Spectroscopic Study of Metallicity, Star Formation, and Electron Density}

\correspondingauthor{John Pharo}
\email{jpharo@aip.de}

\author{John Pharo}
\affiliation{Leibniz-Institut für Astrophysik Potsdam (AIP), An der Sternwarte 16, 14482 Potsdam, Germany}
\affiliation{Department of Physics and Astronomy, University of Missouri, Columbia, MO 65211, USA}

\author{Yicheng Guo}
\affiliation{Department of Physics and Astronomy, University of Missouri, Columbia, MO 65211, USA}


\author{Guillermo Barro Calvo}
\affiliation{University of the Pacific, Stockton, CA, USA}

\author{Teja Teppala}
\affiliation{Department of Physics and Astronomy, University of Missouri, Columbia, MO 65211, USA}

\author{Fuyan Bian}
\affiliation{European Southern Observatory, Chile}

\author{Timothy Carleton}
\affiliation{School of Earth and Space Exploration, Arizona State University, Phoenix, AZ, USA}

\author{Sandra Faber}
\affiliation{UCO/Lick Observatory, Department of Astronomy and Astrophysics, University of California, Santa Cruz, CA, USA}

\author{Puragra Guhathakurta}
\affiliation{University of California, Santa Cruz, CA, USA}


\author{David C. Koo}
\affiliation{UCO/Lick Observatory, Department of Astronomy and Astrophysics, University of California, Santa Cruz, CA, USA}

\begin{abstract}

We present gas-phase metallicity measurements for 583 emission line galaxies at $0.3<z<0.85$, including 388 dwarf galaxies with $\sm < 9.5$, and explore the dependence of the metallicity on the stellar mass and star formation properties of the galaxies. Metallicities are determined through the measurement of emission lines in very deep ($\sim$7 hr exposure) Keck/DEIMOS spectra taken primarily from the HALO7D survey. We measure metallicity with three strong-line calibrations (O3H$\beta$, R23, and O3O2) for the overall sample, as well as with the faint \neiii\ and \auroral\ emission lines for 112 and 17 galaxies where robust detections were possible. We construct mass-metallicity relations (MZR) for each calibration method, finding MZRs consistent with other strong-line results at comparable redshift, as well as with $z\sim0$ galaxies. We quantify the intrinsic scatter in the MZR as a function of mass, finding it increases with lower stellar mass. We also measure a weak but significant correlation between increased MZR scatter and higher specific star formation rate. We find a weak influence of SFR in the fundamental metallicity relation as well, with an SFR coefficient of $\alpha=0.21$. Finally, we use the flux ratios of the \oii\ doublet to calculate gas electron density in $\sim$1000 galaxies with $\sm < 10.5$ as a function of redshift. We measure low electron densities ($n_e\sim25$ cm$^{-3}$) for $z<1$ galaxies, again consistent with $z\approx0$ conditions, but measure higher densities ($n_e\sim100$ cm$^{-3}$) at $z>1$. These results all suggest that there is little evolution in star-forming interstellar medium conditions from $z\sim1$ to $z=0$, confirmed with a more complete sample of low-mass galaxies than has previously been available in this redshift range.

\end{abstract}



\section{Introduction} \label{sec:intro}


Heavy element abundances in the interstellar medium (ISM) of galaxies, or the gas-phase metallicity (Z), is commonly linked to stellar mass, but with considerable observational scatter. The mass-metallicity relation (MZR), wherein the gas-phase metallicity (approximated by the oxygen abundance O/H) is found to increase for galaxies with larger stellar mass, has been well observed in the local Universe with the Sloan Digital Sky Survey \citep[SDSS; e.g.,][]{tremonti2004, zahid2011, am2013}, and much effort has been expended extending the study of this relation to different redshift epochs \citep[e.g.,][]{erb2006,maiolino2008, guo16a, sanders2021}. Even as detection methods have improved and surveys become more thorough, scatter of the order of half an order of magnitude is still observed in measured metallicity at a given stellar mass \citep[e.g.,][]{guo16a}. What drives the metallicity scatter in this relation? Star formation (SF) is a common explanation, as heavy elements produced in stars are dispersed into the ISM, increasing metallicity, while supernova feedback or other gas flow effects may expel enriched gas from the ISM \citep{newman2012, swinbank2019}. The so-called fundamental metallicity relation was developed to connect the star formation rate (SFR) to the stellar mass and metallicity, thereby modifying the MZR to reduce scatter \citep{ellison2008, mannucci2010, cresci2019}, but the metallicity may relate to other properties of the gas as well, such as gas fraction \citep{maiolino_mannucci2019}.

Spectroscopy of nebular emission lines is a technique that has been used to great effect in measuring metallicity and related properties across a wide range of redshift. Most common are strong-line calibrations, whereby ratios of strong forbidden transition metal lines such as \oiii, \oii, \nii, and \neiii, and potentially Balmer emission lines in hydrogen (\ha, \hb) are scaled to metallicity measurements determined via direct measurement (such as with electron temperature derived from the faint \auroral\ emission line) or with comparison to photoionization models \citep[][]{kewley_dopita2002, maiolino2008, curti2017, kewley19}. These calibrations are typically made using star-forming galaxies in the local Universe, but given known offsets in observed emission line ratios, SFRs, and other ISM properties at $z > 1.5$ \citep{steidel2014, strom2017, bian2020}, particular diagnostic calibrations have been constructed to better represent galaxies in different SF epochs \citep{bian2018}. Thus, nebular emission lines may be used to trace the evolution of metallicity for a range of cosmic times and ISM conditions.

To fully grasp the evolution of these metallicity relationships across spans of redshift and stellar mass, it is necessary to also characterize the low-mass galaxy population. Low-mass galaxies, which we typically define as having $\sm < 9.5$, may be more susceptible to events of bursty SF, wherein SF is rapidly triggered and quenched on a timescale of tens of megayears, leading to processes such as expulsion of metal-enriched gas, etc. that can dramatically alter the chemical composition of a galaxy's ISM. Several studies have suggested that bursty SF is more prominent among low-mass or dwarf galaxies \citep[e.g.,][]{searle73,bell01,lee09,meurer09,weisz2012, guo16b}, and theoretical models relating stellar mass and gas-phase metallicity with supernova-driven galactic winds also predict low-mass galaxies will exhibit more scatter in the MZR and FMR \citep{henry2013a, henry2013b,lu15,guo16a}. But it is difficult to obtain comprehensive observations of the dwarf galaxy population at high redshift, since dwarf galaxies are faint at $z>0$. Many higher-redshift studies of the MZR are therefore either limited to the more massive population or probe only those dwarf galaxies with more extreme levels of SF and ionization. This excludes a numerically significant population \citep{muzzin13} of star forming galaxies in a time period where the universe transitions from the cosmic peak of SF \citep{madau14} to the conditions of the local Universe.

In this work, we are able to extend the study of the MZR to dwarf galaxies on the star-forming main sequence \citep{noeske07, whitaker2014} up to $z\sim1$. With deep Keck/DEIMOS spectra from HALO7D, DEEPWinds, and other archival surveys, we measure emission lines for the dwarf galaxies, and therefore are able to determine the gas-phase metallicity via several methods, as well as additional properties including SFR, ionization, and gas density. The mass completeness of the survey also allows for analysis of the intrinsic metallicity scatter as a function of mass and SF. 

The paper is organized as follows. In \S2, we describe the data and sample selection, the emission line flux measurement methods, and appropriate corrections for stellar absorption and dust extinction. In \S3, we describe the methods and calibrations used to measure gas-phase metallicity. \S4 gives the MZR results, and \S5 explores the scatter in the MZR and its possible SF dependency. \S6 describes the measurement of electron densities from \oii\ emission lines, and discusses the implications of our metallicity and density results for ISM conditions in dwarf galaxies. We summarize our findings in \S7.







In this paper, we adopt a flat $\Lambda$CDM cosmology with $ \Omega_m = 0.3$, $\Omega_{\Lambda} = 0.7$, and the Hubble constant $H_0 = 70$ km s$^{-1}$ Mpc$^{-1}$. We use the AB magnitude scale \citep{oke83}.

\section{Data, Sample Selection, and Emission Line Measurement} \label{sec:samp}

\subsection{The HALO7D Survey} \label{sec:h7}

The data used in this paper are comprised of very deep optical spectra of $\sim$2400 galaxies observed with Keck/DEIMOS, and are described in full detail in the HALO7D catalog paper, where the published redshifts, line fluxes, and other measurements may be found \citep{pharo2022}. The spectra were primarily taken by the HALO7D program \citep[PI: Guhathakurta;][]{cunningham2019a}, a program primarily designed to observe faint Milky Way halo stars in the COSMOS, EGS, and GOODS-North CANDELS fields \citep{grogin11, koekemoer11}. Unused space in the DEIMOS slit masks was filled out with galaxies, including a sample of 558 low-mass galaxies at $0 < z < 1.0$ as well as high-mass galaxies targeted for studies of strong winds in star-forming galaxies and stellar populations in quiescent galaxies. Additional programs expand the sample to include GOODS-South, including DEEPwinds (PI: S. Faber), an 8 hr survey yielding $\sim$130 low-mass ($10^8 M_{\odot} < M_{\star} < 10^9 M_{\odot}$) galaxies with F160W AB mag $ < 26.5$; N168D (PI: Livermore), $\sim$70 galaxies; and HALO7D-GOODSS (PI: Kirby), $\sim$100. Dwarf galaxy targets were generally selected to have $0.4 < z < 0.9$, $7.0 < \sm < 9.5$, and F606W mag $<26$. The total observations comprise a sample of 2444 target galaxies, including 1255 low-mass galaxies across four CANDELS fields, as well as 1189 more massive galaxies.

All spectra used in this project were obtained by the DEep Imaging Multi-Object Spectrograph (DEIMOS) instrument at the Keck Observatory \citep{faber03}. The Keck/DEIMOS spectrograph has an overall effective wavelength coverage of roughly $4100 < \lambda < 11000$ \AA. For the HALO7D observations, DEIMOS was configured with the 600 line mm$^{-1}$ grating centered at 7200\AA\, giving a wavelength dispersion of 0.65 \AA\ pix$^{-1}$ (resolution $R \approx 2100$) and a usable wavelength range limited to $5000 < \lambda < 9500$ \AA\ \citep{cunningham2019a}. The slit masks used 1" slit widths, corresponding to 5.4 kpc at $z=0.4$ and 7.9 kpc at $z=0.9$. For $z<1.5$ dwarf galaxies, the 80\%-light radius has been found to be around 5 kpc \citep{mowla2019}, so the slit placement should be sufficient to capture the vast majority of emission from the dwarf galaxies, with relatively little deviation with respect to position angle placement. This radius does expand rapidly as galaxy stellar mass grows beyond $\sm \sim 9$, so slit loss and variation could be more significant in the massive galaxy sample.

The observations were reduced using the automated DEEP2/DEIMOS \textit{spec2d} pipeline developed by the DEEP2 team \citep{newman13}, described fully in \citet{yesuf17} and \citet{cunningham2019a}. This yielded extracted 1D spectra for each exposure, and produced images of the reduced 2D spectra and extraction windows for the purposes of visual inspection of the data. The 2D spectra images were inspected for excessive contamination or other issues, and those exposures that passed visual inspection were co-added into a single 1D spectrum for each galaxy. The co-added spectra were then flux scaled to best-fit photometric spectral energy distributions (SEDs). 

The co-added spectra sample is very deep, with an average combined exposure time of $\sim$7 hr. This represents a substantial increase in observational depth over similar programs such as DEEP2 \citep{newman13}, while also providing a much larger target sample of faint dwarf galaxies than similarly deep contemporary programs, such as LEGA-C \citep{vanderwel2021}. For further details on the co-addition and flux calibration of the spectra or mass/depth statistics of the catalog, see \citet{pharo2022}.

\citet{pharo2022} also conducted several analyses on the properties of the emission line sample relative to comparable CANDELS galaxies. Section 4.2 in that work quantifies the redshift fit success rate as a function of mass, finding similar success fractions ($\sim75\%$) for masses down to $\sm = 8.5$. Section 5.2 of that work places the ELG sample on the star-forming main sequence, finding that the sample does not merely probe starbursting galaxies but instead reaches levels of SF down to the main sequence observed in CANDELS UV photometry \citep{whitaker2014} and below. Color distributions are studied in Section 5.3, finding that the ELG sample is representative of CANDELS galaxies in observed color distributions, but that the low-mass galaxies are bluer in rest-UV colors.

\subsection{Redshift and Emission Line Fitting} \label{sec:z}

\begin{figure*}
    \centering
    \begin{tabular}{c}
        \includegraphics[width=\textwidth]{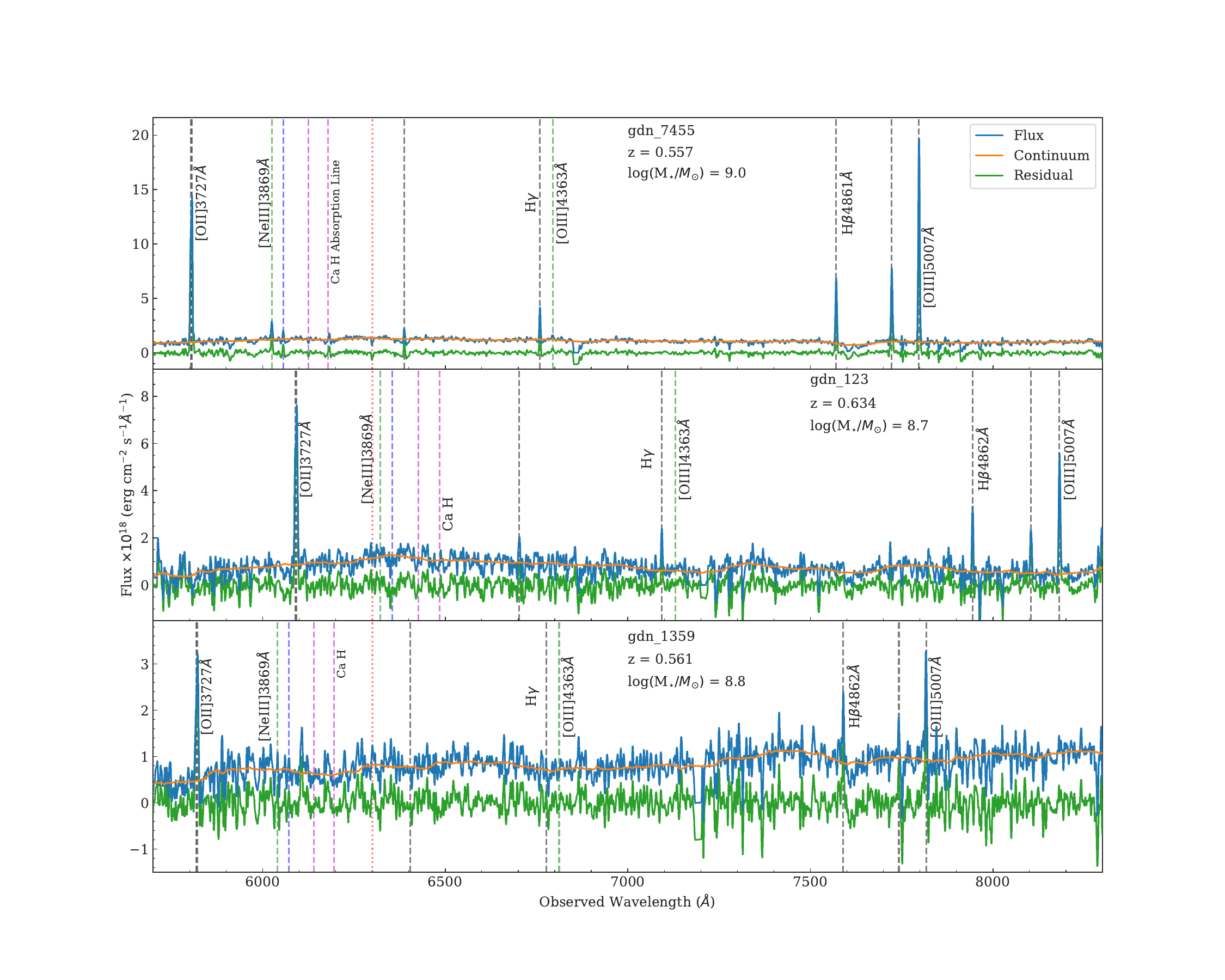} \\
        \includegraphics[width=\textwidth]{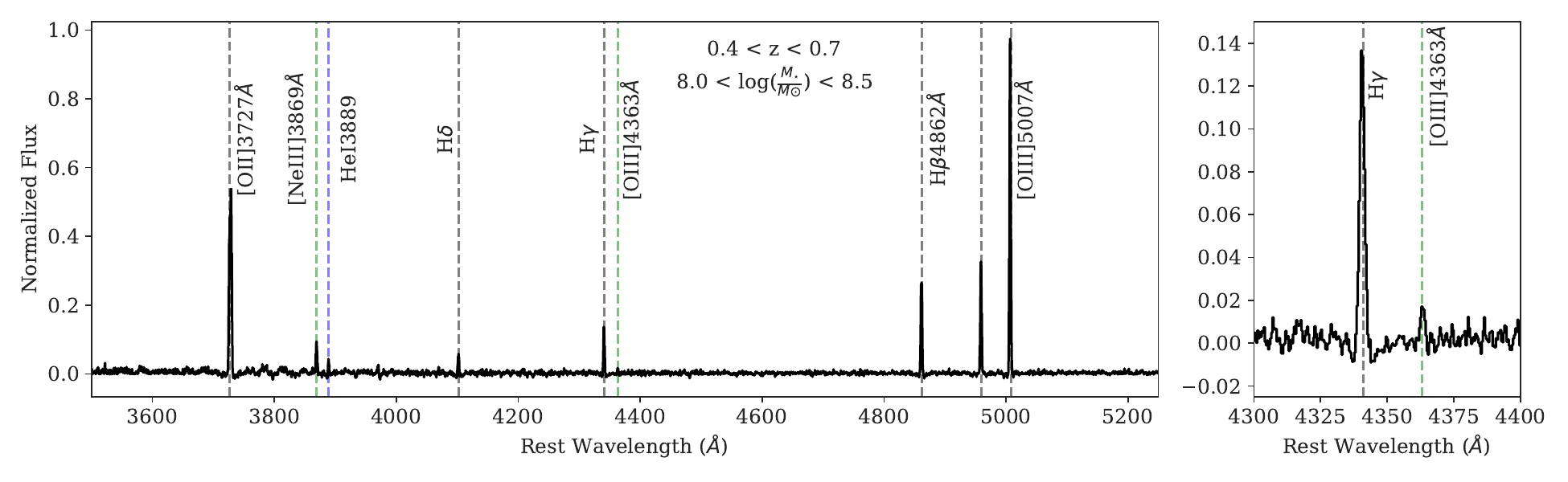}
    \end{tabular}
    \caption{\textit{Top:} example spectra of three HALO7D dwarf galaxies with a range of line EWs. The blue solid line denotes the observed spectrum. The orange solid line shows the continuum estimate, and the green solid line the continuum-subtracted spectrum. Prominent emission lines are labeled and indicated by vertical dashed lines. Black dashed lines indicate strong and/or Balmer series emission lines. Green dashed lines indicate typically fainter ionized metal lines, and blue dashed lines show the locations of faint helium emission lines. The two magenta dashed lines show the locations of the Ca H and K stellar absorption lines, though these features are not prominent in this spectrum, which has little stellar continuum due to its low stellar mass as well as a likely young stellar population. \textit{Bottom:} an example composite spectrum for the lowest-mass bin at $0.4 < z < 0.7$, as described in \S \ref{sec:stacks}. The inset on the right shows a zoom-in around the \auroral\ line used for metallicity measurement via the $T_e$ method.}
    \label{fig:ex_spec}
\end{figure*}

To obtain redshift measurements from the co-added 1D galaxy spectra, we developed a routine to fit strong emission lines in a $0 < z < 2$ redshift window, selected to encapsulate the region where strong-line emitters were likely to be found. The routine expands on the emission line filter technique used in \citet{newman13} and the DEEP2 survey, modified to target emission line galaxies with low or undetectable stellar continua. This was accomplished by fitting a continuum-subtracted spectrum to a redshifted grid of emission and absorption line filters. Prominent spectral lines used in the filters include the H$\alpha$, H$\beta$, H$\gamma$, and H$\delta$ Balmer series lines; the [O \textsc{iii}] and [O \textsc{ii}] ionized oxygen lines; and the Ca H and K absorption lines. H$\alpha$ and the [S \textsc{ii}] doublet are strong enough to include in the fit, but are only present in the handful of $z<0.4$ galaxies included in the target selection. Emission lines that are both faint and rare, such as the [O \textsc{iii}]4363 auroral line, require more careful attention to avoid false detections, and so are measured after redshift fitting is complete. Figure \ref{fig:ex_spec} shows example spectra for HALO7D dwarf emission line galaxies, demonstrating many of these features.


\subsection{Line Flux Measurement and Metallicity Sample Selection} \label{sec:flux}

With the final redshifts determined, we next measured the emission line fluxes, again using methods tested for blind emission line search and low stellar continua \citep{newman13, pirzkal2017, yang17, pharo2019}. First, the continuum flux was estimated and removed throughout the spectrum. For each pixel, a 100-pixel surrounding region is selected. Where possible, this region is selected to include 50 pixels on either side, but this may be redistributed for pixels near the edge of the spectrum. A $3\sigma$ clip is applied to the fluxes in this surrounding region to remove errant skylines or other nearby emission lines. The median of the non-clipped surrounding pixels is taken to be the continuum flux for the central pixel, and is then subtracted from the pixel flux. This process repeats for each pixel in the spectrum. This median subtraction is thus able to account for varying levels of continuum detection, as the median will simply scale toward 0 for low-mass galaxies with low continua, while scaling up to the detected continua of high-mass galaxies. This approach does, however, produce a very smooth continuum estimate that is not well suited to the detection of continuum features. We tested this process using an average of separate continuum measurements for regions red- and blueward of each pixel, and found no substantial offset in final flux measurements. The standard deviation in whatever flux remains from the subtraction in the continuum region is then used to estimate the continuum flux error per pixel, which is combined with the flux measurement error estimates described in \citet{pharo2022}.

Next, the total line flux and flux error is measured by integrating the continuum-subtracted line flux region. This integration is bounded between the points to the left and right of the line center (as defined by the spectroscopic redshift) where the line flux rejoins the continuum, which in the continuum-subtracted spectrum are the points where the residual flux reaches 0. This method avoids potential errors from assuming an incorrect line shape, and is consistent with fluxes derived from lines well fit by a single Gaussian. The ratio of this integrated line flux and the local continuum estimate gives the line equivalent width (EW). Any line for which the ratio of the total line flux to the line error is $\geq3$ is recorded as a detection.

With the redshifts fixed, we include potential faint lines such as [Ne \textsc{iii}]3869, [O \textsc{iii}]4363, and He \textsc{i} and He \textsc{ii} lines. Because these lines are so intrinsically faint, residual skylines or other artifacts in the spectra can easily be mistaken for them. To avoid false positive detections, we first visually screened all spectra, flagging potential detections, and only spectra with a visual flag were fit for faint lines. Because of their utility in measuring gas-phase metallicity and ionization, we will discuss the [Ne \textsc{iii}]3869 and [O \textsc{iii}]4363 detections in this paper. The He line detections will be discussed in subsequent work.

\begin{figure*}
    \centering
    \begin{tabular}{cc}
         \includegraphics[width=0.45\textwidth]{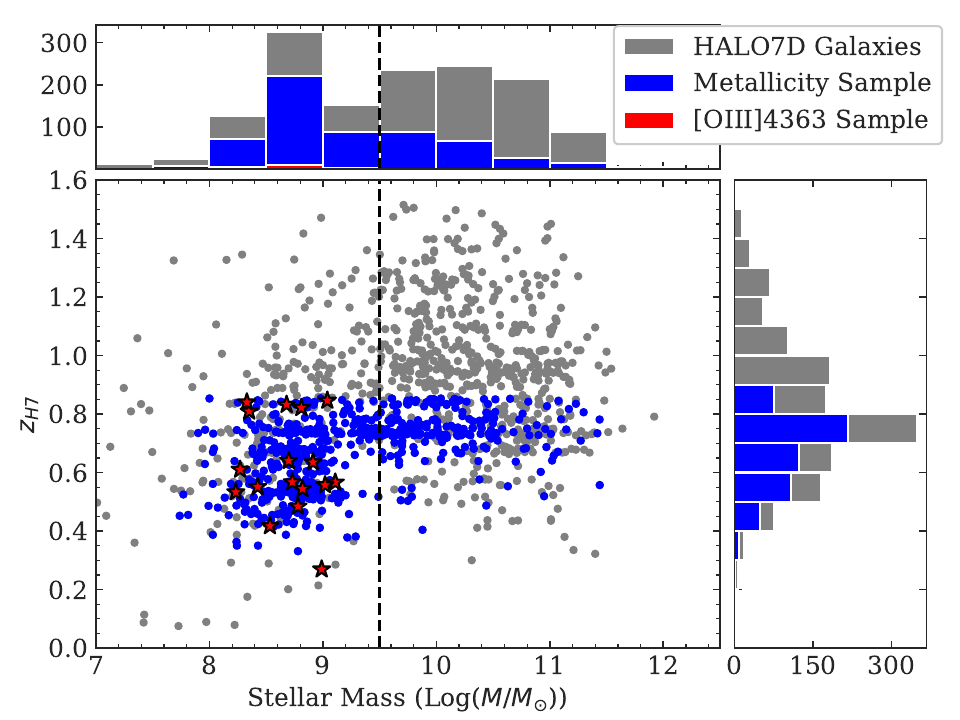} & \includegraphics[width=0.45\textwidth]{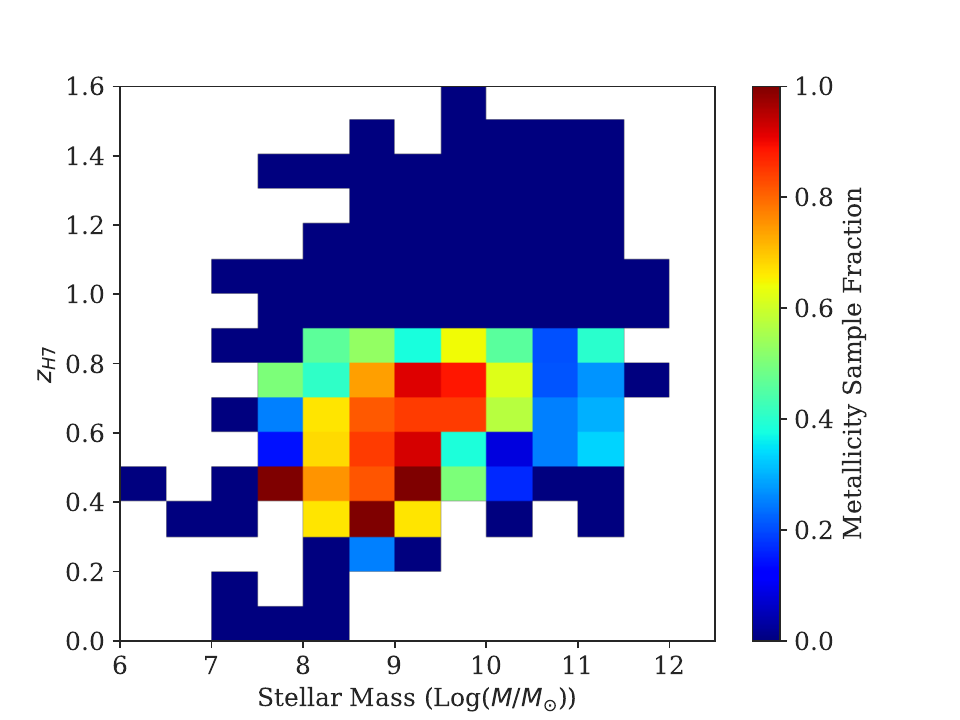}
    \end{tabular}
    \caption{\textit{Left:} the stellar mass versus redshift distribution of the HALO7D galaxy sample. The vertical dashed line separates the dwarf and massive galaxy populations, defined at $\sm = 9.5$. The gray points and histograms indicate the overall sample of galaxies for which a good redshift fit was obtained from spectral features, as described in \S \ref{sec:z}. The blue points and histograms denote the subsample of galaxies with a sufficient set of emission line detections needed to measure the gas-phase metallicity (Z) from one of the strong-line methods described in \ref{sec:metal}. This yields a metallicity sample of 583 galaxies, including 388 dwarf galaxies, with a redshift range of $0.3 < z < 0.9$. At higher redshift, the [O\textsc{iii}]4959,5007 doublet moves out of the detection range. Red stars indicate \auroral\ detections, described in \S \ref{sec:4363}. \textit{Right:} the same distribution, but showing the fraction of the overall HALO7D sample included in the metallicity sample as a function of redshift and mass. The metallicity success rate for dwarf galaxies is high (typically $>75\%$) in the most populated bins. For further analysis of the overall HALO7D sample, see \citep{pharo2022}.}
    \label{fig:mz_dist}
\end{figure*}

From the overall HALO7D emission line catalog, we then selected a sample of galaxies with the requisite emission line detections for measurement of gas-phase metallicity: \oiii, H$\beta$, and \oii.  This effectively imposes a redshift restriction on the metallicity sample of $0.3 < z < 0.9$, limited primarily by the redshifts at which \oiii\ moves off the red end of the DEIMOS detector. With this restriction, we produced a metallicity sample of 583 galaxies, of which 388 are classified as dwarfs. We highlight this sample in blue in the left panel of Figure \ref{fig:mz_dist}.

\subsection{Stellar Absorption and Dust Extinction Corrections} \label{sec:corr}

Hydrogen Balmer emission lines are commonly used in studying the properties of galaxy nebular gas, including SF \citep{kennicutt98}, dust extinction \citep{cardelli1989}, and as part of metallicity calibrations \citep{kewley19}. The lines are also subject to stellar absorption, however, which will reduce the apparent flux if not corrected for. The stellar absorption feature can be measured by fitting a stellar continuum with absorption profiles, but this requires the spectra to observe the continuum with sufficient signal. For our dwarf-dominated galaxy sample, the stellar continua are typically quite low, so instead we adopt corrections from the literature for galaxies of similar mass and redshift. For galaxies with $Log(M_{\star}/M_{\odot}) > 9.5$, we adopt corrections of $EW_{abs}^{H\alpha}=3.4\text{\r{A}}$ and $EW_{abs}^{H\beta}=3.6\text{\r{A}}$ \citep{momcheva2013}, and for galaxies with $Log(M_{\star}/M_{\odot}) \leq 9.5$, we use $EW_{abs}^{H\alpha}=EW_{abs}^{H\beta}=1\text{\r{A}}$ \citep{ly2015}. We add these corrections to the observed Balmer EWs and adjust the measured line fluxes accordingly.

We then used the Balmer lines to measure dust extinction. For each galaxy with both H$\beta$ and $H\gamma$ significantly detected, we calculate the Balmer decrement and the E(B-V) extinction. Then any emission lines are corrected for dust extinction using the E(B - V) measurement and the \citet{cardelli1989} extinction law. For galaxies without both detected Balmer lines, we use the median E(B - V) for galaxies in the same bin of stellar mass to estimate the extinction.

\section{Line Ratios and Metallicity Calibrations} \label{sec:metal}

Many methods exist for measuring the gas-phase metallicity of a galaxy using calibrations of strong-emission line ratios and measurements of faint emission lines. With the depth of HALO7D observations and the width of Keck/DEIMOS wavelength coverage, we are able to assemble large samples of individual line flux measurements for a variety of diagnostic ratios, which we investigate to holistically characterize the emission properties of the HALO7D sample. The individual strong-line flux measurements may be found in Table 4 of \citet{pharo2022}, the HALO7D catalog paper. In this work, we include the derived gas-phase metallicities, line ratios, and new measurements of the \neiii\ and \auroral\ lines in Table \ref{tab:catalog}.

\begin{figure}
    \centering
    \includegraphics[width=0.4\textwidth]{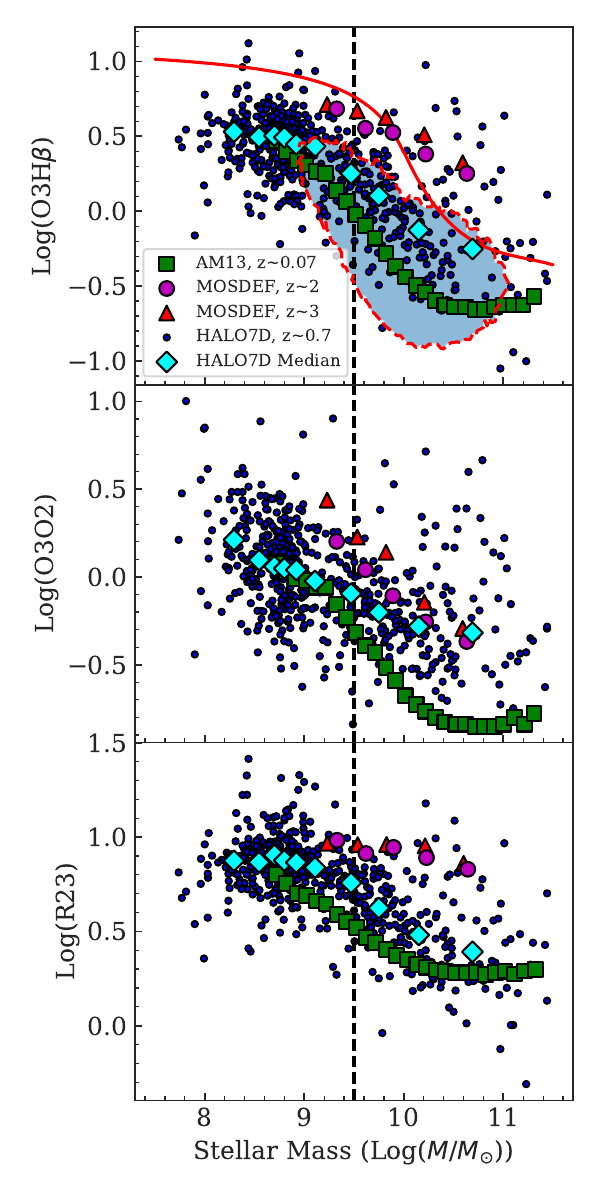}
    \caption{HALO7D strong-emission line ratios as a function of stellar mass. In all panels, blue dots indicate individual HALO7D measurements, and cyan diamonds show the median HALO7D ratios in bins of stellar mass. Green squares denote $z<0.07$ SDSS galaxies \citep{am2013} corrected for diffuse ionized gas emission \citep{sanders2017}. The top panel also gives the 90\% SDSS population distribution within the shaded region. For comparison, composite spectra of $z\sim2.3$ and $z\sim3.3$ star-forming galaxies from the MOSDEF survey \citep{sanders2021} are shown with magenta circles and red triangles, respectively. In the top panel, the red curve indicates the diagnostic cutoff between SF galaxies below and AGN above \citep{juneau2014}. The dashed vertical line separates the dwarf galaxies from the massive sample. In all three methods, the HALO7D medians for dwarf galaxies track the low-redshift composite measurements within the 1 standard deviation scatter, consistent with little evolution in the strong-line ratios of dwarf galaxies from $z\sim1$ to $z=0$. (see \S \ref{sec:strong})}
    \label{fig:ratio_dists}
\end{figure}

\subsection{Strong Lines: O3H$\beta$, O3O2, and R23} \label{sec:strong}

We make measurements of three strong-line ratios for HALO7D galaxies: [O \textsc{iii}]5007/H$\beta$ (hereafter O3H$\beta$)\footnote{Some calibrations of O3H$\beta$ make use of the sum of both the [O\textsc{iii}4959 and 5007 emission lines. Since the ratio of the 4959 and 5007 lines is fixed to $\sim$3, however, it is trivial to move between such calibrations and the 5007-only version preferred here. In this text, O3H$\beta$ refers to [O \textsc{iii}]5007/H$\beta$ unless otherwise indicated.} [O \textsc{iii}]4959,5007/[O \textsc{ii}]3727,3729 (O3O2), and (O \textsc{iii}]4959,5007 + [O \textsc{ii}]3727,3729)/H$\beta$ (R23).

Each ratio has benefits and drawbacks as metallicity indicators. O3H$\beta$, comprised of two lines relatively close in wavelength, is not as sensitive to the effects of dust extinction as ratios composed of lines farther apart, reducing the impact of a common major source of uncertainty. Since this ratio does not include flux contributions from the singly ionized oxygen, though, it is also dependent on the ionization parameter and temperature of the system in addition to the gas-phase metallicity. O3H$\beta$ calibrations are also double branched, meaning a low value of the ratio can correspond to either high or low metallicity, requiring some further information to break the degeneracy. 

O3O2 is a monotonic calibration, so does not require a degeneracy breaker, and is not dependent on Balmer emission lines that are also used to trace SF. Its constituent lines are far enough apart to require correction for dust extinction, though, and the calibration of the ratio to the metallicity has a larger dispersion than other methods \citep{maiolino2008}. R23 incorporates the most emission line information, and so is less dependent on other factors such as the ionization parameter, but does require both dust correction and a mechanism for breaking branch degeneracy.

Figure \ref{fig:ratio_dists} shows the distributions of these strong-line ratios in HALO7D as a function of galaxy stellar mass, sometimes known as mass-excitation diagrams. In the O3H$\beta$ distribution in the top panel, we include the diagnostic curve from \citet{juneau2014}. Galaxies that lie above this curve on the excitation diagram are likely hosts to active galactic nuclei (AGN), while those below are excited purely by SF. We note, however, that this diagnostic has not been thoroughly tested for dwarf galaxies at $z>0$. For the HALO7D galaxies above the diagnostic curve, we attempted to confirm AGN activity through other tests. Searches of IRAC infrared and Chandra X-ray databases yielded no indicators of AGN activity for matching dwarfs, and the dwarfs lack near-infrared spectra that would enable use of the BPT diagnostic \citep{bpt1981}. Nevertheless, we flag all HALO7D galaxies above the \citet{juneau2014} diagnostic curve as potential AGN, and so exclude them from the subsequent metallicity analyses. For the low-mass galaxies where AGN activity is less certain, the flagged sample includes six galaxies, a small fraction of the overall dwarf sample, so the analysis is unlikely to be significantly impacted.

To study redshift evolution in the emission ratios, we also include distributions from studies at lower and higher redshifts. For $z\sim0$, we use stacks of emission line galaxies in bins of mass from SDSS \citep[][often AM13]{am2013}, which have been corrected for contributions from diffuse ionized gas (DIG) so that the ratios represent solely H II region emission \citep{sanders2017}. The fractional contribution of DIG emission is expected to be highest for galaxies with low SF surface density, and so is unlikely to be a significant contribution to dwarf galaxies at higher redshift, which are more strongly star forming \citep{henry2021, pharo2022}. We therefore do not apply a DIG correction. We also compare with stacks of ELGs at $z\sim2.3$ and $z\sim3.3$ from the MOSDEF survey \citep{sanders2021}, so as to better place the HALO7D sample within cosmic redshift evolution.

Figure \ref{fig:ratio_dists} shows that the dwarf galaxy population in HALO7D largely tracks the $z\sim0$ dwarf galaxies, though this depends on the diagnostic used. For galaxies with $\sm < 9.5$, the median offsets between the HALO7D O3H$\beta$ measurements and $z\sim0$ sample is 0.05 dex, well within the typical standard deviation of 0.2 dex in each HALO7D bin. The O3O2 ratio is similar, with a median offset of -0.01 dex and scatter of 0.23. R23 shows a more noticeable distinction, with a median offset of 0.1 dex, though this is still within the 1 standard deviation HALO7D scatter of 0.16 dex. All diagnostics show larger offsets from the high-$z$ comparisons, ranging from 0.15 dex in R23 to 0.47 in O3O2, though each high-$z$ sample has just one dwarf-mass bin to compare with. On the whole, the strong-line diagnostic measurements from $z\sim0.7$ HALO7D dwarfs are more consistent with those observed at $z=0$.

\subsection{[Ne \textsc{iii}]3869\AA\ and Ne3O2} \label{sec:neon}

With the significant depth of the HALO7D observations, we do not need to rely solely on strong-line methods to study metallicity and ISM conditions. For many galaxies, measurements of intrinsically fainter emission lines are available in our sample. The catalog in \citet{pharo2022} records significant detections of the H$\gamma$ and H$\delta$ Balmer emission lines as well as the stronger and more common H$\beta$. After these lines, the most commonly detected fainter line is \neiii. 

The HALO7D \neon\ sample has been analyzed in detail in \citet{pharo2023}, which reported 167 \neon\ detections, including 112 in dwarf galaxies. This enables the study of the log(\neiii/\oii) ratio (Ne3O2) distribution as a function of stellar mass, akin to the distributions shown in Figure \ref{fig:ratio_dists}. That paper demonstrates that the typical \neon\ emitter in HALO7D tracks the low-redshift SDSS galaxies, with composite spectra measurements showing a median offset of 0.0 dex with AM13. This corroborates the results of the strong-line emitters in this work, and we adopt the \neon\ measurements for a fourth method of measuring the metallicity. Due to the intrinsic faintness of the line, however, these individual \neon\ detections likely represent some of the strongest \neon\ emitters at $z\sim0.7$, and so exist between the composite measurements at $z\sim0$ and z$\sim2$.

While the \neon\ sample is smaller and potentially less representative because of this, it does provide some advantages. As with O3H$\beta$, the close wavelength proximity of \neiii\ and \oii\ make Ne3O2 a diagnostic much less susceptible to uncertainty driven by dust extinction. And as with O3O2, it has a metallicity calibration that is monotonic \citep{maiolino2008}, eliminating major errors in metallicity measurement due to branch confusion. We therefore include Ne3O2 metallicity measurements in our analysis.

    

\subsection{[O \textsc{iii}]4363\AA\ and $T_e$ Metallicities} \label{sec:4363}

The aforementioned metallicity measurements rely on calibrations of emission line ratios to either photoionization modeling or methods of directly measuring the metallicity via the electron temperature. These calibrations necessarily have some scatter in relation to other metallicity measures, as well as the other potential drawbacks described in the sections above. The ratio of the \auroral\ emission line to the \oiii\ lines is sensitive to the electron temperature, and therefore provides a measure of heavy element abundance in ionized gas for a given value of electron density. Observation of this ratio then provides a ``direct'' measure of the gas-phase metallicity, rather than a calibration. This has made observation of the auroral line a common method for measuring metallicity in local H II regions and nearby galaxies \citep[e.g.,][]{izotov2006}, as well as providing new calibrations for strong-line methods \citep[][]{bian2018,jiang2019}.

The downside of such a method is that the auroral \auroral\ line is intrinsically very faint, typically producing flux lower than \oiii\ by a factor of $\sim$100 in all but the most ionized systems. Consequently, the line is rarely observed, and samples in which it is detected are likely not representative of the overall line-emitting galaxy population. Nevertheless, several previous studies have attempted to directly measure the metallicity in galaxies at $z\sim0.7$ in order to provide a check on strong-line methods of measurement \citep{ly2014,jones2015,pharo2019}. With the deep observations from HALO7D and focus on dwarf galaxies, we search for \auroral\ line emission in order to expand the sample of direct metallicity measurements at this redshift.

In order to identify \auroral\ emission lines in the HALO7D spectra, we limited our sample to the dwarf galaxies. We checked for significant emission in two ways: by visual inspection, and through a modified application of the automated line-fitting routine used in \citet{pharo2022}. Galaxies where \auroral\ emission lines were both flagged in visual inspection by three individual observers and where the line was detected with a signal-to-noise ratio $S/N > 3$ in the line-fitting routine were selected for the \auroral\ sample. This yielded 21 significant detections of the \auroral\ emission line.

We measured the electron temperature ($T_e$) from the emission lines in order to determine the metallicity following the method described in \citet{ly2014}, also described in \citet{pharo2019}. This method requires detection of the \oiii, \oii, and H$\beta$ emission lines as well, which does reduce the size of our usable sample to 17 galaxies. Though small, this is still an increase in the sample size of dwarf galaxies with $T_e$-based metallicity measurements at this redshift. However, this detected sample represents just a small fraction of the total HALO7D dwarf galaxy sample ($\sim4\%$), and may not depict the typical metallicity and ionization conditions for dwarf galaxies at $z\sim0.7$. To gain a better understanding of the median properties of star-forming dwarfs, and to make use of our whole dataset, we constructed composite spectra, described in the following section.

\begin{deluxetable*}{cccccccccccccc}
\centering
\label{tab:catalog}
\tablecaption{Metallicity Catalog}
\tablecolumns{14}
\tablehead{
\colhead{ID} &
\colhead{$z_{H7}$} &
\colhead{Mass} & 
\colhead{Z(O3H$\beta$)} & 
\colhead{Z(R23)} &
\colhead{Z(O3O2)} &
\colhead{Z(Ne3O2)} &
\colhead{Z($T_e$)} &
\colhead{O3H$\beta$} &
\colhead{eO3H$\beta$} &
\colhead{R23} &
\colhead{eR23} &
\colhead{O3O2} &
\colhead{eO3O2} 
}
\startdata
cos-10050 & 0.5175 & 9.489 & 8.98 & 8.79 & 9.06 & -99 & -99 & 0.44 & 0.17 & 0.14 & 0.13 & 4.64 & 0.07 \\
cos-1022 & 0.6826 & 10.79 & 7.44 & 7.47 & 7.84 & -99 & -99 & 3.67 & 0.88 & 4.63 & 0.09 & 5.97 & 0.81 \\
cos-1029 & 0.7282 & 9.42 & 8.53 & 8.67 & 8.42 & -99 & -99 & 2.33 & 0.14 & 1.15 & 0.05 & 5.79 & 0.13 \\
cos-10659 & 0.6046 & 8.969 & 8.5 & 8.47 & 8.53 & -99 & -99 & 2.51 & 0.08 & 0.84 & 0.06 & 7.34 & 0.07 \\
cos-10774 & 0.7313 & 11.01 & 8.24 & 8.29 & 8.29 & -99 & -99 & 4.34 & 0.23 & 1.67 & 0.06 & 9.27 & 0.2 \\
cos-10976 & 0.736 & 9.405 & 8.7 & 8.62 & 8.76 & -99 & -99 & 1.38 & 0.08 & 0.42 & 0.06 & 6.23 & 0.06 \\
\enddata
\tablenotetext{}{Table containing redshifts, stellar masses, gas-phase metallicities, and emission line ratios for each diagnostic, dust-corrected fluxes for \neiii\ and \auroral\ line detections, and excitation diagnostic flags (columns truncated for space). The full table is available in its entirety in machine-readable form.}
\end{deluxetable*}

\subsection{Composite Spectra} \label{sec:stacks}

\begin{deluxetable*}{cccccccc}
    \label{tab:stack}
    \centering
    \tablecaption{Composite Spectra Bins}
    \tablecolumns{8}
    \tablehead{
    \colhead{$z_{min}$} &
    \colhead{$z_{max}$} &
    \colhead{Median $z$} &
    \colhead{Median Mass} &
    \colhead{Mass Range} &
    \colhead{N} &
    \colhead{$Z_{4363}$} &
    \colhead{$Z_{O3H\beta}$}}
    \startdata
    0.4 & 0.7 & 0.56 & 8.38 & 8.0-8.5 & 62 & 8.10 & 8.35 \\
    0.4 & 0.7 & 0.57 & 8.77 & 8.5-9.0 & 155 & 8.14 & 8.44 \\
    0.4 & 0.7 & 0.56 & 9.11 & 9.0-9.5 & 20 & 8.30 & 8.48  \\
    0.7 & 0.9 & 0.77 & 8.38 & 8.0-8.5 & 28 & 8.09 & 8.26 \\
    0.7 & 0.9 & 0.77 & 8.79 & 8.5-9.0 & 76 & 8.19 & 8.38 \\
    0.7 & 0.9 & 0.76 & 9.35 & 9.0-9.5 & 86 & 8.26 & 8.50 \\
    \enddata
    
\end{deluxetable*}

The analysis of emission line measurements in individual galaxies is necessarily limited to those galaxies whose observations have the signal necessary to detect the emission lines. This introduces possible selection biases, in particular for samples of intrinsically faint emission lines, such as \auroral. By combining groups of individual spectra into composite spectra, we may obtain average measurements for subsamples of galaxies that include galaxies without individual detections of a given emission line. This may then provide a more representative measure of the emission characteristics of that galaxy subsample.

For this stacking procedure, we limit our sample to dwarf galaxies with $Log(M_{\star}/M_{\odot}) < 9.5$ so as to avoid the mixing of the quiescent/AGN/wind candidates in the higher mass samples (see \citet{pharo2022, tacchella2022, wang2022}). Within the dwarf sample, we further select galaxies with a significant [O \textsc{iii}]4959,5007\AA\AA\ detection in \citet{pharo2022}, and with a redshift fit of $0.4 < z < 0.9$. This should not bias the sample, since [O \textsc{iii}]4959,5007\AA\AA\ is typically the brightest emission line detected, and the redshift restriction ensures that all of the emission lines necessary for metallicity and SF analysis will be covered in each individual spectrum going into the composite. Stellar masses are taken from catalogs in \citet{santini15} and \citet{barro19}.

After the sample is selected, we sort the dwarf galaxies by redshift and mass into bins containing comparable numbers of galaxies. In order to produce bins with enough constituent galaxies to yield a meaningful average, we use three bins in stellar mass and two in redshift. The details on the bin sizes and their constituent galaxies are described in Table \ref{tab:stack}. 

For each bin, we then combine the individual spectra with the following procedure. First, each spectrum is corrected for dust extinction using the \citet{cardelli1989} extinction law and using the measurement as described in \S \ref{sec:corr}. Then the continuum is estimated and subtracted, and the residual fluxes are normalized to the [O \textsc{iii}]5007\AA\ line flux. We choose to initially remove the continuum from all galaxies in order to avoid difficulties with particularly low-mass galaxies, where the continuum is often not well detected. Normalization to the [O \textsc{iii}]5007\AA\ flux eliminates any issue of relative flux dimming from the slightly different redshifts among galaxies in the same bin, and since we are primarily concerned with emission line ratios rather than their fluxes, we may operate with this normalization. Furthermore, as one of the consistently brightest emission lines in virtually all galaxies in the sample, this normalization helps to ensure that the highest SFR galaxies do not overly influence the composite measurement.

Next, the normalized spectrum is rebinned onto a uniform grid of wavelengths. Once each spectrum in the redshift-mass bin has been normalized and rebinned, they may be stacked together by taking the median flux at each wavelength. Normalized emission line fluxes may then be measured from each composite spectrum, along with the median stellar absorption as a fraction of line emission. For lines with possibly significant stellar absorption, we measure the flux and absorption by simultaneously fitting emission and absorption profiles. To obtain errors for the emission line measurements, we use a Monte Carlo bootstrap method wherein the constituent galaxies of the bin are resampled with replacement, and the new samples stacked and emission lines measured. The errors for each emission line are then estimated from the standard deviation in the resulting distribution of measurements. See the bottom panel of Figure \ref{fig:ex_spec} for an example composite spectrum.

For faint sources, it is possible that median stacking introduces a bias through systematic underestimation of brightness, which may impact fainter emission lines more significantly \citep{steidelhamilton1993}. To check for this, we perform an alternative calculation of the composite spectra using the mean flux rather than the median, and then recalculate the emission line ratios and resulting metallicities. We find no distinction in the method for the strong-line ratios, where all the lines are well-detected. The mean stacks do yield somewhat larger \auroral\ fluxes, resulting in lower $T_e$ measurements of Z, but the differences are consistently less than 0.1 dex across mass and redshift bins, within the composite spectra errors, so this potential bias does not influence our result. 

Table \ref{tab:stack} describes the redshift and stellar mass bins of the composite spectra. For the dwarf galaxies, we separated the spectra into $8.0 < \sm < 8.5$, $8.5 < \sm < 9.0$, and $9.0 < \sm < 9.5$ bins of stellar mass. We chose fixed widths of the stellar mass bins to preserve a range of stellar mass in the final composite spectra. In order to have a sufficient number of galaxy spectra contributing to each bin, we limited the composite spectra to two redshift bins: $0.4 < z < 0.7$ and $0.7 < z < 0.9$. This maximizes the signal gained through stacking while preserving the possibility of detecting any redshift evolution in the MZR from $z=1$ to $z < 0.4$. 

\begin{figure*}
    \centering
    \includegraphics[width=\textwidth]{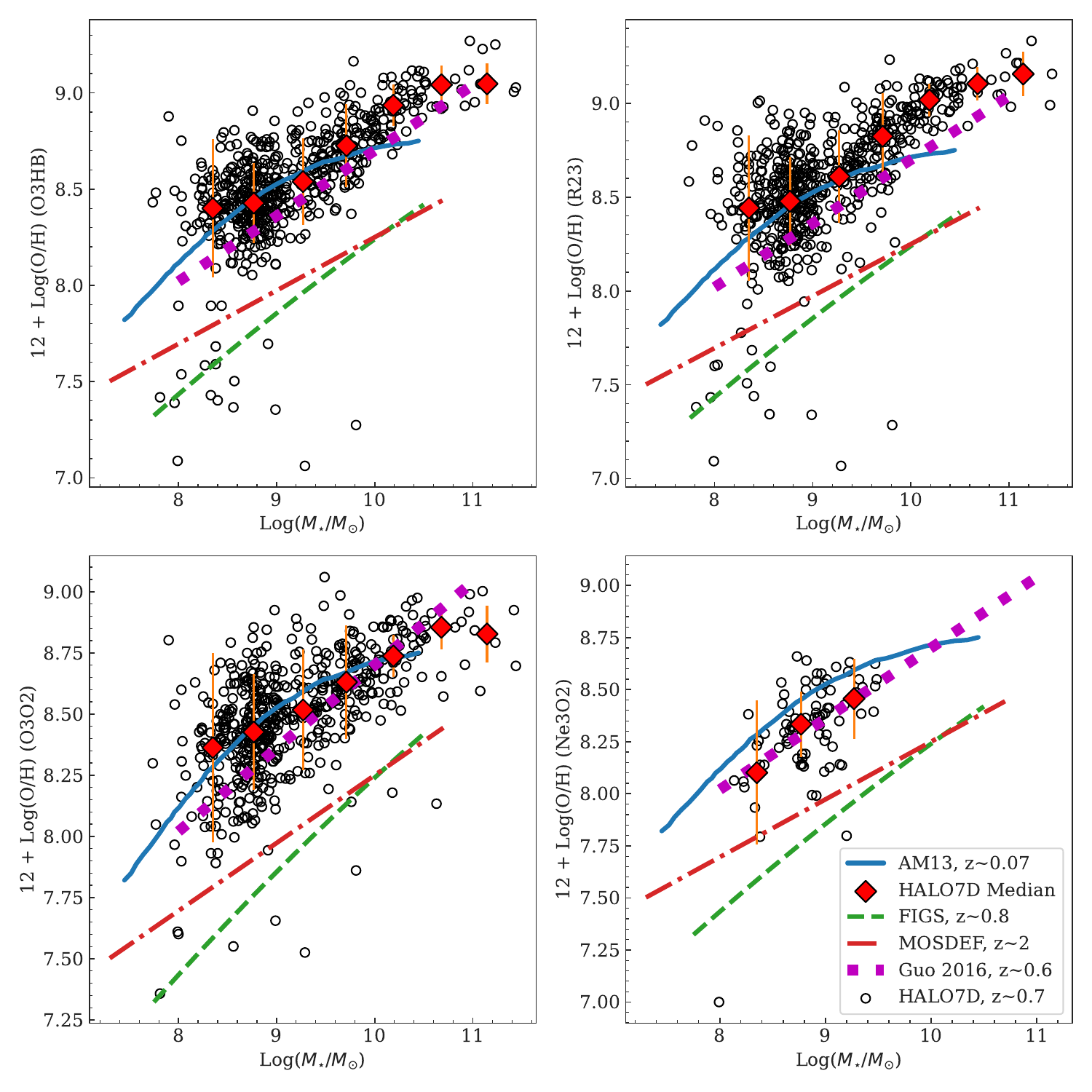}
    \caption{The HALO7D MZRs. Each panel depicts an MZR determined with a different metallicity indicator. Clockwise from top left, they use O3H$\beta$, R23, Ne3O2, and O3O2. Individual HALO7D galaxies are shown as hollow black circles. The median HALO7D metallicity in bins of stellar mass is shown by the red diamonds, and the error bars give the standard deviation of metallicity in each mass bin. The HALO7D MZRs given by the median metallicities are consistent within the 1$\sigma$ scatter with MZRs measured at redshifts near 0 \citep{tremonti2004, am2013} and those measured at comparable redshift using strong-line methods \citep{zahid2011, guo16a}. This means we see a high-metallicity offset in comparison the MZRs measured at higher redshift \citep[e.g.,][]{sanders2020}, as well as those determined using only $T_e$-measured galaxies dependent on the \auroral\ line. We also see some indication of higher scatter in metallicity among low-mass galaxies, which we explore in the subsequent sections.}
    \label{fig:mzr}
\end{figure*}

\section{The Mass-Metallicity Relation at $z\sim0.7$} \label{sec:mzr}

We use each of the line ratios described in Section \S \ref{sec:metal} to measure gas-phase metallicities for the HALO7D galaxies. For O3H$\beta$, O3O2, R23, and Ne3O2, we use the diagnostics derived in \citet{maiolino2008}, which combines calibrations of strong-line ratios to $T_e$-derived metallicities at low metallicity with calibrations to photoionization models at higher metallicity. These calibrations use the oxygen abundance $12 + log(O/H)$ as a proxy for the overall metallicity, and for convenience, we will refer to this abundance as the metallicity $Z$. 

The line ratio-metallicity distributions are then fit to a polynomial. Best-fit calibrations for O3O2 and Ne3O2 are monotonic in the observed metallicity range of $7 < Z < 9.5$, so for these we simply apply the polynomial function to our observed ratios to obtain the metallicity. The O3H$\beta$ and R23 fits are double branched, however, with low ratio values potentially corresponding to either very high or very low Z. To break this degeneracy, we follow \citet{guo16a} and use the O3O2 ratio. \citet{henry2013a} demonstrated that the O3O2 ratio is an effective means of identifying low-Z branch galaxies, and following the O3O2 calibration from \citet{maiolino2008}, we adopt \oiii/\oii$>3$ as a cutoff for using the the lower metallicity branch fit for our O3H$\beta$ and R23 observations. 

We obtained stellar mass measurements from SED-fitting catalogs from CANDELS \citep{santini15,barro19}, COSMOS/Ultravista \citep{muzzin13}, and EGS/IRAC \citep{barro11a, barro11b}. We then construct MZRs for the metallicity sample for each of the strong-line methods and Ne3O2. The relations are shown in Figure \ref{fig:mzr}, with individual HALO7D galaxies marked with black circles. We measured the median metallicity in bins of mass, shown with red diamonds. The standard deviation of metallicity measurements in each bin is given by the error bars. 

We produced the MZR for the full mass range of the metallicity sample for the three strong-line methods, but for Ne3O2, we restricted the metallicity measurement to the dwarf galaxy sample to avoid additional uncertainty due to measuring low-EW lines in galaxies with strong stellar continua. We note as well that there are large systematic offsets in comparing metallicities derived from different strong-line calibrations \citep{kewley2008}, potentially up to 0.8 dex. Therefore, the most direct comparison will be between HALO7D $Z_{O3H\beta}$ and similarly derived metallicities from \citet{guo16a}.

The MZRs derived from all four metallicity indicators show the expected trend of increasing metallicity with increasing stellar mass. The three strong-line methods also show the flattening of the relation at high mass that is commonly seen at low redshift \citep{tremonti2004,zahid2014}. For comparison, we have included a non-comprehensive sample of MZRs measured in other surveys at a range of redshifts. We measure relations that are consistent within the 1$\sigma$ HALO7D scatter with both strong-line method measurements at comparable redshift \citep[e.g.,][]{guo16a} and with those at lower redshift \citep{am2013}. However, the HALO7D MZRs have higher median metallicity at given stellar mass compared to \citet{guo16a} by 0.1-0.2 dex, particularly with the R23 method, which reaches a 0.3 dex offset at higher masses. This difference is smaller than the size of the standard deviation of the HALO7D measurements for the dwarf sample, but could potentially also be due to the deep HALO7D observations capturing more of the population of higher metallicity weak line emitters. Note that the Ne3O2 sample, representing some of the strongest emitters in the metallicity sample, tracks the \citet{guo16a} result with no detectable offset.

We include as well the MZR measured at $z\sim2.2$ in the MOSDEF survey via the $T_e$ method. As with most MZR studies around the $z\sim2$ peak in cosmic SF, they observe a low-metallicity offset at fixed stellar mass compared to the MZR at lower redshift, in this case approximately 0.5 dex lower than the HALO7D medians. The magnitude of this offset depends at least somewhat on methods of selection and metallicity measurement, however. In \citet{sanders2021}, MOSDEF spectra are stacked into composite spectra, and metallicities measured from the stacks from strong-line ratios. At comparable redshift, the MOSDEF stacks yield metallicity measurements only 0.2 dex lower than the HALO7D MZR. The $z\sim0.8$ MZR measured from FIGS \citep{pharo2019} via the $T_e$ method is substantially offset from HALO7D at comparable redshift as well, ranging from 0.5 dex at $\sm \approx 9.5$ in O3H$\beta$ to 0.8 dex at the lowest masses. This highlights a difficulty in measuring metallicity at higher redshifts: the direct method of measurement with \auroral\ is typically only available for the most metal-poor sub-population of galaxies. 

\begin{figure*}
    \centering
    \includegraphics[width=\textwidth]{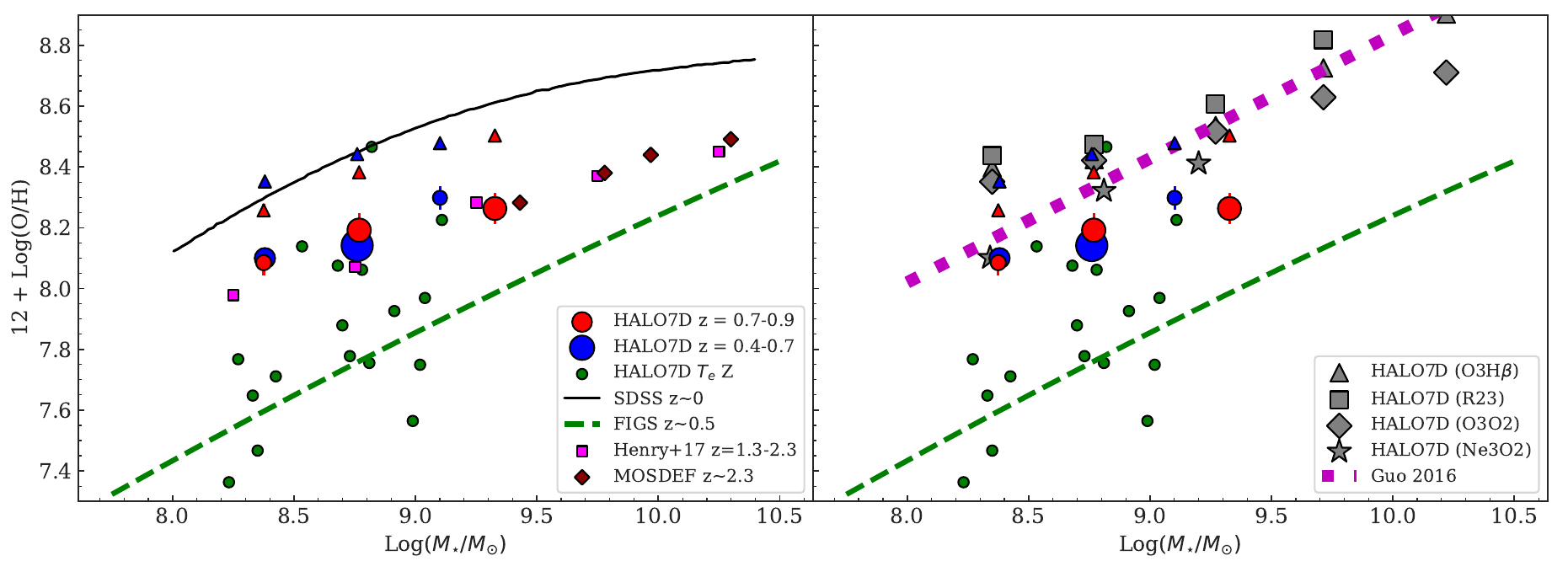}
    \caption{The MZR using $T_e$-derived metallicity measurements. Individual HALO7D galaxies are shown as small green circles. Blue and red circles show the measurements from composite spectra of HALO7D galaxies in two redshift bins, with the points sized according to the number of spectra contributing to each composite. Blue and red triangles show O3H$\beta$-derived metallicities for the same stacks. In the left panel, we include for comparison a $z\sim0$ MZR from $T_e$ measurements of SDSS galaxies \citep{am2013}, $T_e$-derived metallicities from a small sample of $z\sim0.8$ galaxies \citep[FIGS,][]{pharo2019}, $1.3 < z < 2.3$ stacks \citep{henry2021}, and $z\sim2.3$ stacks \citep{sanders2021}. In the right panel, we compare with strong-line calibrations in both HALO7D and from \citet{guo16a}.}
    \label{fig:mzr_4363}
\end{figure*}

To address this, we also construct a HALO7D MZR for individual galaxies with a significant \auroral\ detection, and with $T_e$-based metallicities derived from the stacked spectra. This MZR is shown in Figure \ref{fig:mzr_4363}. The most direct comparison for these measurements is with the FIGS sample, which is at similar redshift and also determined through the \auroral\ line. The HALO7D sample has a narrower mass range, but a larger number of dwarf galaxies. We find a comparable MZR, except with a larger scatter due to the measurement of several relatively high-Z \auroral\ galaxies with $Z>8.0$. 

Blue and red circles give the measurements of the stacks at $z\sim0.56$ and $z\sim0.77$, sized according to the number of galaxies contributing to each stack. The stack metallicities are noticeably higher than the median individual metallicity measurement (0.5 dex at the lowest mass, 0.3 at the highest), which is comparable to the FIGS MZR. This discrepancy confirms that the \auroral-detected subsample is lower metallicity than the typical $z\sim0.7$ line emitter. Blue and red triangles show the O3H$\beta$ metallicities measured from the stacks, where we see a similar 0.2 dex offset from the $T_e$ measurement as described above in MOSDEF. 

The possible source of this offset is important, as the $T_e$-based metallicities from the composite spectra show very little evolution from composite measurements at $1 < z < 2.5$ \citep{henry2021,sanders2021}, obtained from Bayesian fitting methods and strong-line calibrations, respectively. The median offset between a linear fit of the stack metallicities and linear fits to the $1 < z < 2.3$ ($z\sim2.3$) stacks is 0.075 (0.06) for the lower redshift stack and 0.055 (0.04) for the higher redshift spectrum, compared to offsets of 0.35 and 0.33 from O3H$\beta$. This would imply that the evolution in harder ionizing conditions at $z\sim2$ to $z=0$ \citep{strom2017} is still largely ongoing at $z\sim0.7$, which is not what we find with our strong-line calibrations. 

One possible explanation is that the $T_e$ method underestimates the global metallicity in a galaxy by overemphasizing hotter H II regions within the galaxy over cooler regions where \auroral\ is weaker, resulting in offsets in measured metallicity of up to 0.4 dex \citep{stasinska2005, bresolin2007}. If the offset seen between HALO7D stack $T_e$ measurements and the same measurements at $z\sim0$ \citep{am2013} can be explained as a result of the HALO7D stacks underestimating contributions from high-metallicity galaxies, then these results could still be consistent with the strong-line observations in indicating little to no metallicity evolution in dwarf galaxies from $z\sim0.8$ to $z=0$.

The size of this offset is also within the metallicity variation attributable to different choices of strong-line calibration, however. Offsets between different strong-line methods can differ from each other by $>0.5$ dex \citep{kewley19}, even when using the same line ratio. Some calibrations are designed specifically to emulate the conditions of $z\sim2$ galaxies in lower-$z$ analogs \citep{bian2018}, which can yield higher Z measurements for a given O3H$\beta$ values compared to more generic calibrations, but such calibrations are designed for O3H$\beta$ ratios higher than those observed in the HALO7D stack sample. If this offset is the result of inadequate strong-line calibrations for metallicity in this mass-redshift regime, then the direct metallicities indicate only a fraction of the ISM evolution from $z>1.5$ to $z=0$ has occurred by $z\sim0.7$. It may require further study of line ratio diagnostics at spatially resolved scales at non-local redshift to help resolve this discrepancy.




\section{Star Formation, Metallicity Scatter, and the Fundamental Metallicity Relation} \label{sec:met_and_sfr}

\subsection{Star Formation Rates} \label{sec:sfr}

Our detection of hydrogen Balmer emission lines enables the estimation of the SFR from the line flux. These hydrogen recombination lines are driven by ionizing radiation from massive, short-lived stars, and therefore trace the $<10$ Myr or instantaneous SFR. We calculate the SFR for HALO7D galaxies using the \citet{kennicutt98} formulation where $\text{SFR } = 7.9\times10^{-42}L_{H\alpha}$ and assuming Case B intrinsic flux ratios of $f_{Hn}/F_{H\alpha}$. SFR may also be calculated from calibrations of the [O \textsc{ii}] luminosity, but this is less precise as L([O \textsc{ii}]) is sensitive to other parameters, including the metallicity. Therefore, we preferentially use Balmer lines to determine the SFR, preferring the most intrinsically strong line detected. Full analysis of the SFR properties and the star-forming main sequence of HALO7D galaxies may be found in \citet{pharo2022}, which demonstrated that the deep HALO7D observations probe the dwarf galaxy population down to sub-main-sequence SFRs in emission lines. Consequently, our analysis is not limited to the starbursting dwarf population, but instead spans the range of SF properties in dwarf galaxies.

\subsection{Metallicity Scatter} \label{:sec:Z_scatter}

Our large sample of individual metallicity measurements enables us to make a more comprehensive analysis of the scatter in gas-phase metallicity on the MZR, especially among lower mass galaxies where previous studies are less complete at $z\sim0.7$. In particular, we are interested in exploring the properties of galaxies that may relate to or influence the scatter, in order to better understand what drives differential galaxy metallicity at a given redshift.

\begin{figure*}
    \centering
    \includegraphics[width=\textwidth]{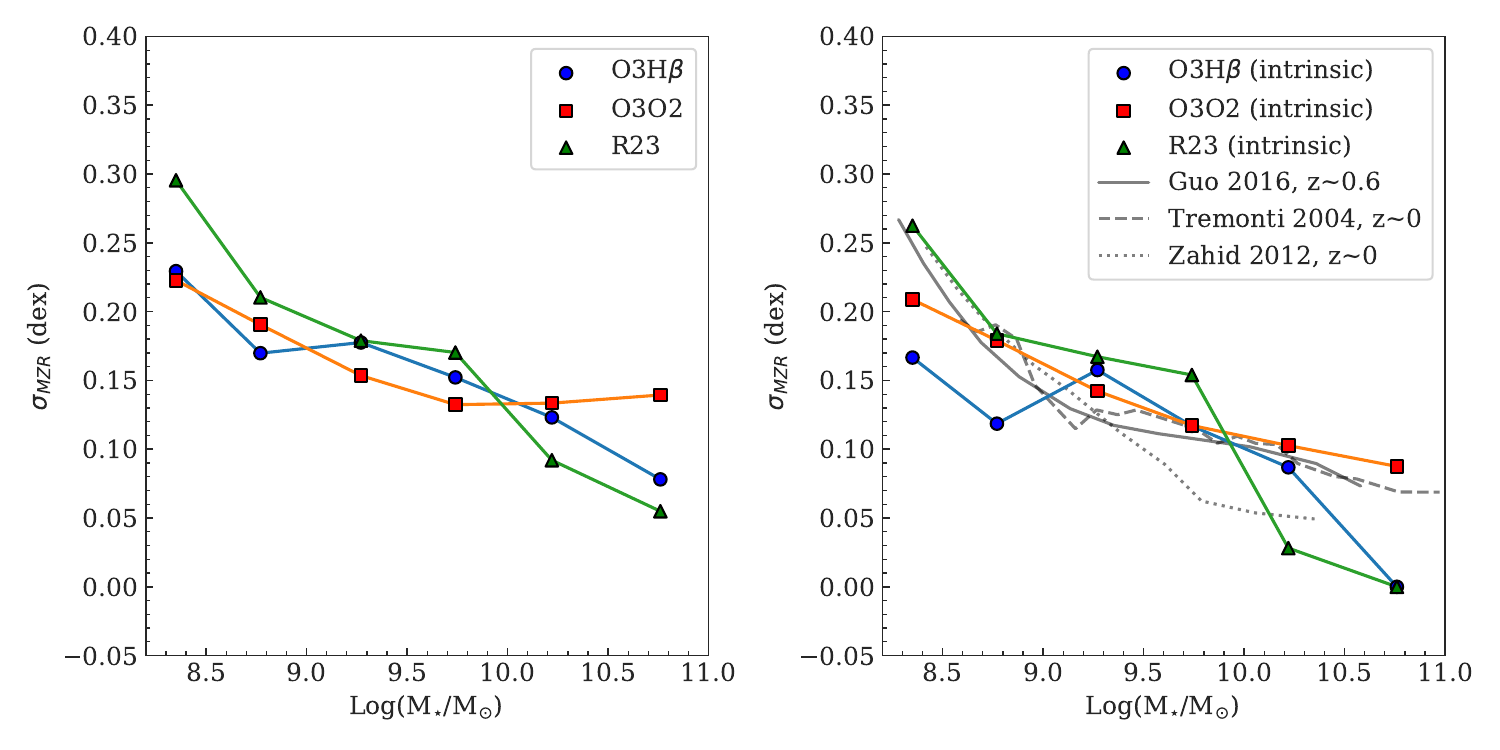}
    \caption{\textit{Left:} the scatter in the MZR $\sigma_{MZR}$ in bins of stellar mass. We calculate $\sigma_{MZR}$ for each of the three strong-line metallicity methods, indicated by different shapes and colors of track. \textit{Right:} the intrinsic $\sigma_{MZR}$ scatter, calculated by subtracting in quadrature the median measurement error in each bin from the measured scatter. The trend of increasing scatter remains after removal of measurement error, indicating that a broader range of gas-phase metallicities is fundamentally related to lower stellar mass in a galaxy. Gray curves indicate comparison samples, which show similar overall trends and offsets of $<0.05$ dex for low-mass galaxies.}
    \label{fig:mass_scatter}
\end{figure*}

To check metallicity scatter as a function of mass, we bin the galaxies by stellar mass and measure the difference between the 16th and 84th percentile metallicities in the bin, called $\sigma_{MZR}$. We perform this calculation for each of the three strong-line methods, and show the results in the left panel of Figure \ref{fig:mass_scatter}. We find a general increase in $\sigma_{MZR}$ as stellar mass decreases for all three methods, though with some variation. Scatter in O3H$\beta$ and O3O2 increases by only about 0.2 dex from the most massive bin to the least, while the $\sigma_{MZR}$ in R23 increases by 0.4 dex.

At this point, we want to distinguish between scatter driven by flux measurement error, which could plausibly change as a result of observational differences of spectra from galaxies of different stellar mass, from the intrinsic scatter of the metallicities for galaxies in a mass bin. To obtain the intrinsic $\sigma_{MZR}$, we subtract in quadrature the median measurement error in each bin. We show the intrinsic scatter as a function of stellar mass in the right panel of Figure \ref{fig:mass_scatter}. The difference between the observed and intrinsic scatter typically amounts to 0.05 to 0.1 dex per bin, with the notable exception of the most massive bins, where measurement error seems to be the significant driver of the scatter, particularly for the O3H$\beta$ and R23 methods. The large measurement error scatter in the massive sample is potentially related to the stronger stellar continua in the more massive galaxies, leading to larger stellar absorption of Balmer emission lines affecting the H$\beta$ line measurement. This would explain the strong impact on the two methods using the H$\beta$ line, while the O3O2 measurements are less affected. The lower line EWs in the massive galaxy measurements may also lead to greater error from continuum estimation and subtraction.

Given the already discussed nonuniform selection of the massive galaxies, it is safer to consider the change in intrinsic scatter from the 10.5 mass bin down to lower mass. The right panel of Figure \ref{fig:mass_scatter} shows that there is still a significant increase in scatter as the mass decreases. This largely corroborates previous measurement of the intrinsic $\sigma_{MZR}$ at this redshift in \citet{guo16a}, who measured an increase of $\sigma_{MZR}=0.1$ at $\sm = 10.5$ to $\sigma_{MZR}=0.3$ at $\sm \sim 8-8.5$ using just the O3H$\beta$ method of metallicity measurement. This new measurement increases the dwarf galaxy sample size by a factor of $\sim$2. In \citet{guo16a}, the increased scatter at low mass was driven primarily by a long tail of very low-metallicity dwarf galaxies. While such low-Z galaxies are observed in this analysis as well (see top left of Figure \ref{fig:mzr}, the somewhat shallower slope we find for the MZR indicates increased detection of high-Z low-mass galaxies in HALO7D, likely a result of deeper observations yielding detections of galaxies with fainter line emission.

The trend in scatter observed here and in \citet{guo16a} closely matches the trend observed in low-mass galaxies at $z=0$. \citet{guo16a} measured intrinsic $\sigma_{MZR}$ from the MZR measurements \citet{tremonti2004} and \citet{zahid2012} made of large samples of $z=0$ SDSS and DEEP2 galaxies. We present these measurements here, and see the same anti-correlation with mass observed in HALO7D with small offsets (almost always $<0.05$ dex) for low-mass galaxies. The exception is O3H$\beta$, where we measure a shallower anti-correlation and therefore find an offset of $\sim0.1$ dex in the lowest mass bin. However, we note that the curve presented by \citet{guo16a} is the best fit of several metallicity diagnostics, subsamples of which show substantial offsets from each other in the low-mass regime (see Figure 7 in that paper). Diagnostic offsets aside, the consistency of the anti-correlation across samples and methods indicates no substantial redshift evolution in the MZR scatter of dwarf galaxies from $z\sim0.7$ to $z=0$.

\begin{figure*}
    \centering
    \includegraphics[width=\textwidth]{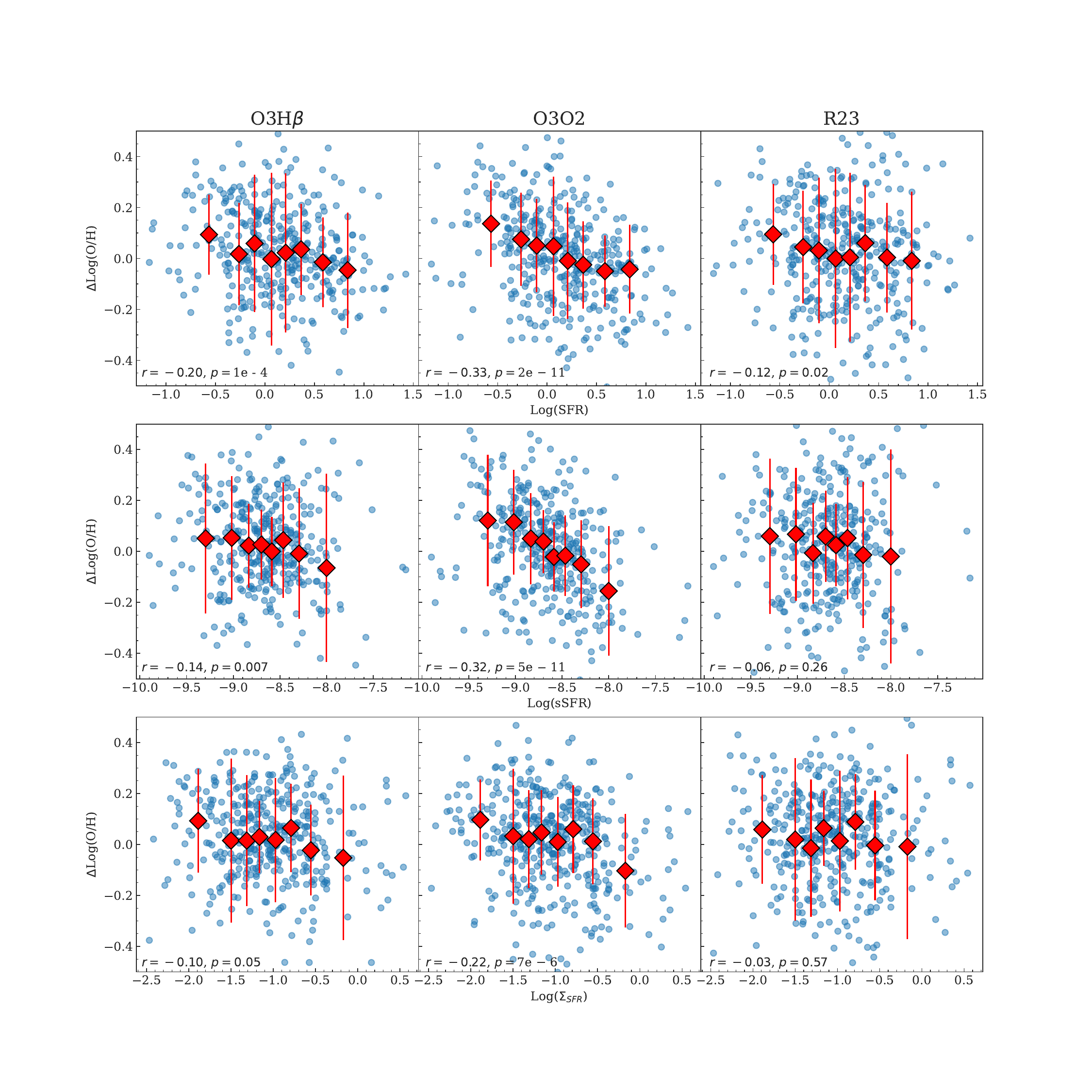}
    \caption{\textit{Top:} each panel gives $\Delta \log(O/H)$ as a function of log(SFR) for $Z_{O3H\beta}$ (left), $Z_{O3O2}$ (middle), and $Z_{R23}$ (right). Individual HALO7D galaxies are given by blue points, while median $\Delta \log(O/H)$ is given by red diamonds for bins of log(SFR). Error bars give the standard deviation in each bin. Using Spearman rank correlation tests, we find weak but significant correlations between $\Delta \log(O/H)$ and the SFR, though the significance for R23 is marginal. The test parameters are shown in the bottom left of each panel. \textit{Middle:} same as the top row, but $\Delta \log(O/H)$ is given as a function of sSFR, the SF per stellar mass. We test this to check whether the scatter dependence on SFR is not simply restating the stellar mass dependence. We once more find weak but significant negative correlations with O3H$\beta$- and O3O2-measured metallicities, but no significance with R23. The methodological dependence perhaps reflects the use of the H$\beta$ line in both Z and SFR measurements for O3H$\beta$ and R23, but with O3O2, we measure significant correlations with both SFR and sSFR, indicating a weak tendency for more star-forming galaxies to exhibit lower Z. \textit{Bottom:} $\Delta \log(O/H)$ is given as a function of $Log(\Sigma_{SFR}$), the SF surface density, calculated from the half-light radii in kiloparsecs. We test this to check whether morphology helps SF relate to the metallicity scatter, e.g. by facilitating or inhibiting the expulsion of metal-enriched gas. As in the two previous cases, we measure a weak, marginal anti-correlation using O3H$\beta$, a significant anti-correlation with O3O2, and no detectable relation with R23.}
    \label{fig:sfr_scatter}
\end{figure*}


\subsection{Metallicity Scatter and Star Formation Dependence} \label{sec:sfr_scat}

Whether a galaxy's placement on the MZR has a dependence on SF has been a common subject of analysis for metallicity at a range of redshifts. At low redshift, several studies have found that at given stellar mass, galaxies with higher SFR tend to have lower metallicity \citep{ellison2008, mannucci2010, cresci2019}, though some recent studies have questioned this \citep{sanchez2019}. Evidence of a mass-Z-SFR relation has been found up to redshifts of $\sim3.5$ \citep{hunt2016, sanders2021}, but the strength of the relation and whether it evolves with changing redshift remains in question \citep[e.g.,][]{henry2021}. Given this and the increase in MZR scatter measured in HALO7D galaxies, we next investigate possible SFR dependence for HALO7D metallicities.

Initially, we looked for direct correlations in the MZR scatter with the SFR. For this analysis, we limit the HALO7D sample to just the dwarf galaxies, in order to avoid issues arising from the nonuniformity of the massive galaxy selection. Focusing on just the dwarf galaxy MZR also avoids any need to account for flattening in the MZR shape. 

For each strong-line metallicity measure, we do a simple linear fit to the dwarf galaxy mass regime of the $\sm$-$log(O/H)$ relation, and then define a \textit{true} MZR function based on this fit. Then for each galaxy measurement, we calculate the difference between the measured and true metallicity values, $\Delta \log(O/H) = Z_{obs} - Z_{mod}$. Figure \ref{fig:sfr_scatter} shows $\Delta \log(O/H)$ calculated for each of the three strong-line methods as a function of Log(SFR), Log(sSFR), and Log($\Sigma_{SFR}$). 

For each panel in the figure, we perform a Spearman rank correlation test to see if significant correlations exist between metallicity offset and SF. This test gives two numbers. First, the correlation coefficient, which has a range of values $-1 < r < 1$, with 1 representing the strongest positive correlation between two variables, and -1 the strongest negative correlation. Second, it gives the significance $p$. The results of the tests are given in each figure panel. We find significant ($p<0.05$) but relatively weak negative correlations between SFR and $\Delta$Log(O/H) for all three methods ($-0.12 < r < -0.33$ for all diagnostics), though only marginally with R23. However, since SFR correlates with stellar mass, and we already established a correlation with scatter and stellar mass, we need to check for SF dependence in isolation. 

For this, we compare with the specific SFR (sSFR), the SFR per stellar mass. Here we still find significant but weak correlations with the O3H$\beta$ and O3O2 methods, but the significance in the R23 correlation disappears. We do find a stronger and more significant result with O3O2 compared to the two other methods, though still only with $r=-0.32$. This is the only method not subject to branch confusion and not dependent on measurement of the H$\beta$ line, which we also use to measure the SFR. Finally, we check as well the scatter's dependence on $\Sigma_{SFR}$, the SF surface density. We calculate this with the previously measured SFRs and galaxy half-light radii measured in Galfit using near-infrared CANDELS photometry \citep{vanderwel2012}. This yields a very similar result to that of the previous tests.

On the whole, these results point to a weak dependence of metallicity on the SF, such that higher SF makes a galaxy more likely to exhibit lower metallicity. This is similar to the trend in MZR residuals with SFR observed in low-$z$ dwarf galaxies \citep{cresci2019} as well as in more massive SF galaxies at $z\sim2.3$ \citep{sanders2020}.

\subsection{The Fundamental Metallicity Relation} \label{sec:FMR}

Next, we test for SFR's ability to reduce MZR scatter by measuring an FMR for the HALO7D metallicity sample. As described in \S\ref{sec:intro}, the FMR has been used to reduce MZR scatter by plotting Z versus $\mu = \log(M_{\star}/M_{\odot}) - \alpha \log(SFR)$, where $\alpha$ is a coefficient representing the strength of the SFR dependence. Measurement of $\alpha$ varies from study to study, possibly due to redshift evolution or selection effects.

To test the FMR in our sample, we calculate the RMS scatter of the relation $Z \propto M_{\star} - \alpha \cdot \log_{10}(SFR)$ for values of $\alpha$ from 0 to 1 in steps of 0.01. We find the RMS minimized with an $\alpha$ of approximately 0.21 for all three strong-line methods, though the reduction in scatter is very small, changing the RMS scatter from 0.25 to 0.24. This corresponds to the findings of \citet{guo16a}, which indicate only a weak sSFR dependence at similar redshift, and \citet{henry2021}, who measure a similarly small change in RMS with a best-fit $\alpha=0.17$ for $z\sim1-2$ galaxies, though the overall level of scatter measured in HALO7D is $\sim0.05$ dex higher, perhaps due to deeper observation of low-SF dwarfs. This is a much weaker effect than what is sometimes measured for the local SFR-MZR dependence ($\alpha=0.32$ in \citet{mannucci2010}, $\alpha=0.66$ in \citet{am2013}). 

It has been suggested that this could be a distinction between strong-line and \auroral\ methods for measuring metallicity, with the latter more correlated with changing SFR \citep{am2013, sanders2017, henry2021}. Checking the HALO7D \auroral\ sample, we find no significant relationship between our $T_e$ metallicities and SFR or sSFR, and \citet{pharo2019} found only a weak correlation in galaxies at similar redshift, though both samples are quite small compared to local studies. This also does not necessarily agree with what has been found at $z\sim2$ \citep[e.g.,][]{sanders2020}. \citet{curti2023} explore preliminary FMR results from $z>3$ dwarfs from JWST, and find that the local FMR parameterizations with high $\alpha$ do not effectively describe this sample. 

The exact relation of local SF to the gas-phase metallicity is thus difficult to determine, and may depend upon intrinsic biases and scatter in the metallicity calibrations used, which can be quite large. Future observations with JWST that expand the sample of \auroral\ emitters at higher $z$ may help address this uncertainty. However, that we are able to confirm weak SF dependence in the dwarf galaxy population indicates the importance of deep observations probing down to less-active SF galaxies in order to not be biased substantially in favor of the most active star-forming galaxies.

\section{Electron Density and Discussion} \label{sec:edens}

\begin{figure*}
    \centering
    \includegraphics[width=\textwidth]{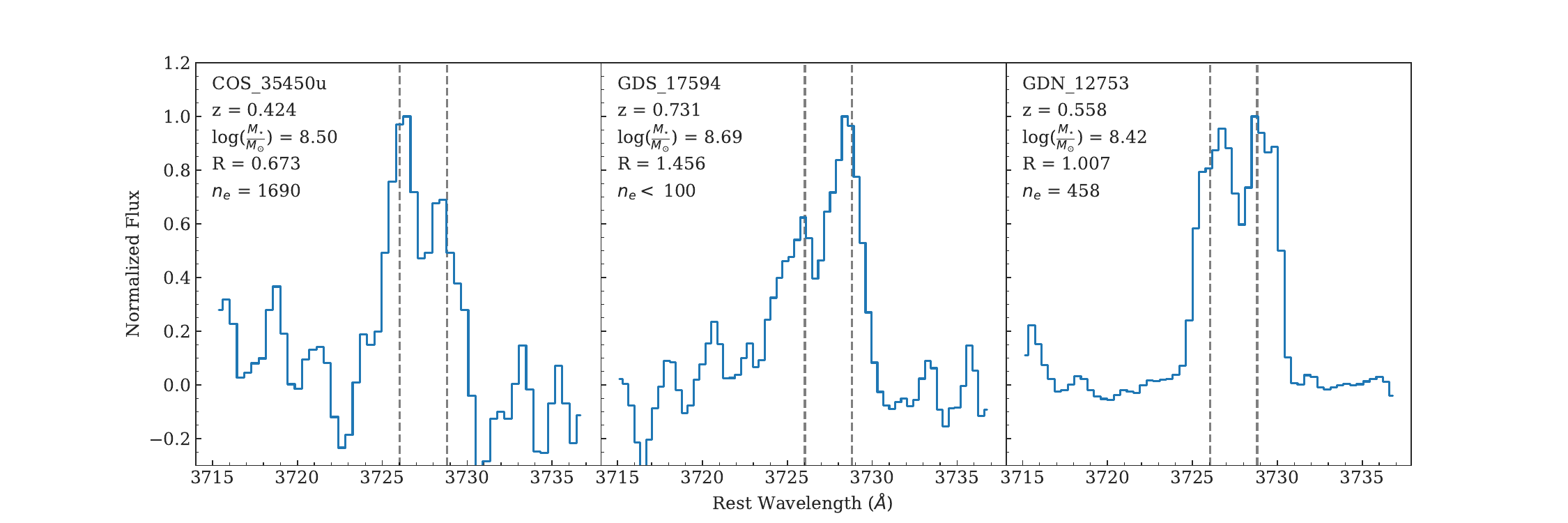}
    \caption{The \oii\ doublet for three different HALO7D dwarf galaxies showcasing a range of line ratios and electron densities. The continuum-subtracted spectrum normalized to the peak line flux is given in blue, with the in-atmosphere rest wavelengths of \oii\ given by the dashed vertical lines. The inset text gives the HALO7D ID, emission-line-derived redshift, stellar mass, \oii\ doublet ratio $R$, and electron density $n_e$ in cm$^{-3}$ for each galaxy. The density for GDS\_17954 is reported as an upper limit since $R$ becomes insensitive to the electron density for $n_e \lesssim 100$ cm$^{-3}$.}
    \label{fig:o2_ex}
\end{figure*}

The \oii\ doublet is a commonly used tracer of the gas electron density in the H II regions of star-forming galaxies. The two doublet lines have a small difference in excitation potential which, at typical H II region temperatures, makes the lines sensitive to collisional excitation and de-excitation rates. Collisional rates are strongly dependent on the electron density of the ionized gas, so the ratio $RO2$ of the fluxes of these two lines is sensitive to the density \citep{kewley19}. This provides an observational tracer for the electron density that is readily available given sufficient spectral resolution.

With a line separation of $\sim$2.5 \AA, the \oii\ doublet line peaks may be resolved, but the profiles of the two lines will be semi-blended in the HALO7D spectra. To obtain accurate flux measurements for each line, we perform a simultaneous fit of two gaussian profiles, with the ratio of the wavelength centroids fixed to the known \oii\ wavelength ratio within the value of one DEIMOS spectral element. We also require that the widths of the two profiles match, and fix the range of allowable flux ratios between the two lines to $0.3839 < RO2 < 1.4558$, corresponding to limiting electron densities of 100,000 and 1 cm$^{-3}$ from the models in \citet{sanders2016}, respectively. For this procedure, we use the \textit{lmfit} Python fitting package \citep{newville_matthew_2014_11813}. See Figure \ref{fig:o2_ex} for example spectra.

With $RO2$ calculated from the fits of the \oii\ emitters, we could then measure electron densities from a calibration of $n_e$ to $RO2$. For this, we use the calibration developed in \citet{sanders2016}. The major uncertainty in the calibration arises from the temperature dependence of the collision strengths; \citet[][hereafter S16]{sanders2016} develop the calibration on the assumption of HII regions that are neither particularly metal-rich nor metal-poor. Given the metallicity distribution of the HALO7D sample (Figure \ref{fig:mzr}) shows a sample with high scatter that is not skewed to high or low metallicity relative comparable samples, we expect the uncertainty from metallicity-driven temperature variations to be negligible compared to the intrinsic scatter of the sample. We do note that the S16 calibration assumes an electron temperature of 10,000 K, and consequently will underestimate the electron density of any high-temperature dwarfs by up to 20\%. However, the small size of the $T_e$-detected sample suggests this will not dominate the whole dwarf sample.

The S16 electron density calibration for \oii\ is given by

\begin{equation}
    n_e(RO2) = \frac{638.4RO2 - 930.7}{0.3771 - RO2}
\end{equation}

\begin{figure}
    \centering
    \includegraphics[width=0.45\textwidth]{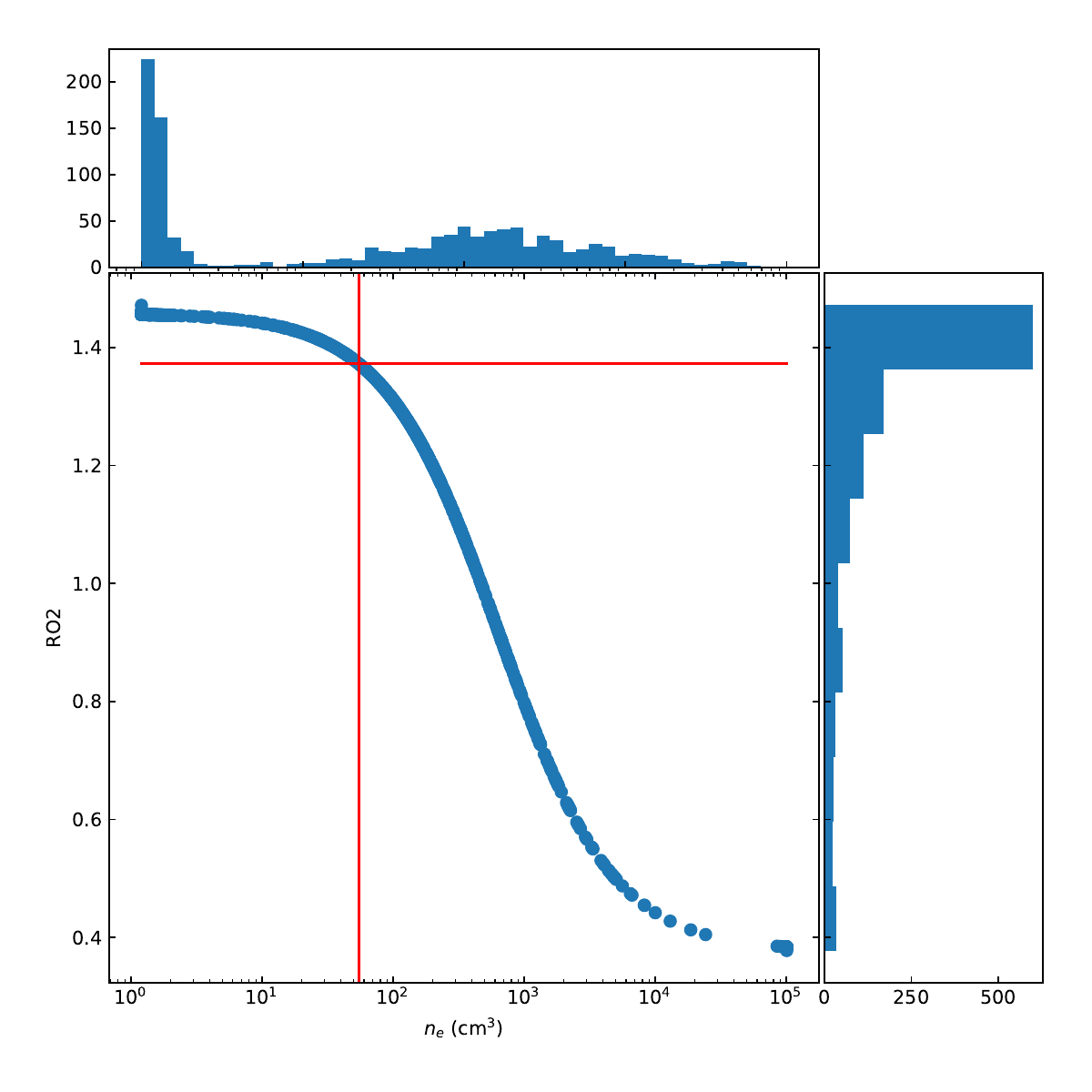}
    \caption{The $RO2$-$n_e$ distribution for HALO7D \oii\ emitters. Red lines give the median $R=1.37$ and $n_e=55$ cm$^{-3}$ values for the sample. However, the histograms of the $R$ and $n_e$ distributions show that this is far from a uniform distribution: nearly half the sample is clustered at very low $n_e$. This is largely due to the insensitivity of the $R$-$n_e$ calibration for densities below $\sim50$ cm$^{-3}$, where the changes in $R$ are much smaller than the measurement error.}
    \label{fig:R_ne}
\end{figure}

\noindent The resulting $RO2$-$n_e$ distribution for HALO7D \oii\ emitters is shown in Figure \ref{fig:R_ne}. The median values of $RO2$ and $n_e$ for the sample are 1.37 and 55 cm$^{-3}$, respectively, but the individual values are clearly not uniformly distributed throughout the allowable ranges. As can be seen in the histograms, nearly half the sample is clustered at very low electron density (high $RO2$). This is largely a result of the insensitivity of the ratio calibration in the low-density regime, where a very small range of $RO2$ values span nearly two orders of magnitude in density.

\begin{figure}
    \centering
    \includegraphics[width=0.45\textwidth]{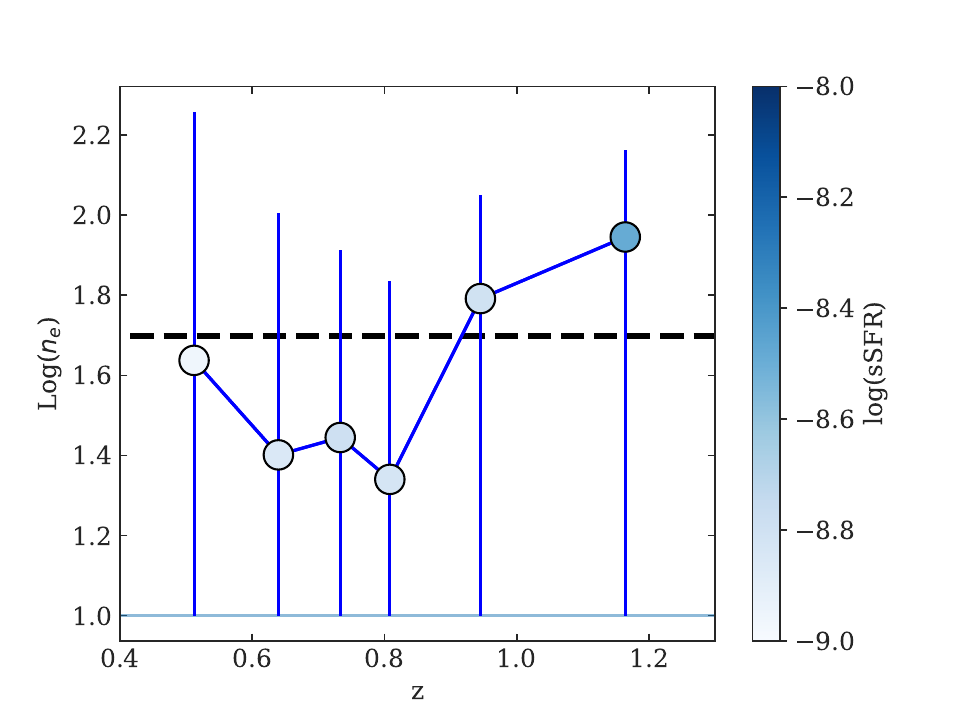}
    \caption{The electron density as a function of redshift for HALO7D dwarf galaxies. Each point shows median $n_e$ in a redshift bin of 162 galaxies, with error bars giving the interquartile range of $log(n_e)$ for the bin. The shading of the points denotes the median sSFR for each bin. The solid horizontal line indicates the low-density limit from \citet{sanders2016} for measuring $n_e$ from the \oii\ doublet ratio, and the dashed horizontal line gives the value at which the density becomes insensitive to the flux ratio. This shows that the dwarf galaxies at $z<1$ have low electron densities consistent with those typical of $z\sim0$ galaxies, while at $z>1$, the dwarf galaxy sample has higher densities as well as elevated sSFR, though the scatter in density is high throughout.}
    \label{fig:ne_z}
\end{figure}

If we limit the distribution to only those galaxies with $\sm < 10.5$, in order to exclude the nonuniformly selected massive galaxy sample, the shape of the distribution remains very similar. The median $RO2$ changes slightly to 1.39, though this does shift the median density down to 40 cm$^{-3}$. To explore possible evolution with redshift, we plot the median electron density in bins of redshift in Figure \ref{fig:ne_z}, focusing on the dwarf galaxy sample.

For $z<0.9$ dwarf galaxies, there is only superficial variation in the median density, as the medians are typically at or below the $RO2$ sensitivity limit of $n_e\sim50$ cm$^{-3}$. For $z>0.9$, however, median densities for dwarf galaxies are appreciably higher and above the sensitivity limit, reaching $n_e \sim 100$ in the highest redshift bin ($z\sim1.2$). Scatter in the electron density, shown by the interquartile range, is high in all bins. The SF conditions shift as well, with median sSFR increasing with redshift. 

To place our low-mass galaxies in context, we compare with the summary of previous results in \citet{isobe2023}, which collects a number of studies of electron densities in the ISM over a redshift range of $0<z<9$, including some recent high-redshift findings with JWST. The four $0<z<1$ studies \citep{berg2012, kaasinen2017, swinbank2019, davies2021} measure the 16th-84th percentile range of $n_e$ to be 25-100 cm$^{-3}$, consistent with our findings in HALO7D. However, only one of these studies \citep{berg2012} included a significant sample of galaxies with $\sm < 9$, and these consisted of only local, $z\sim0$ galaxies. The other three have median masses $\sm > 9.5$, above our cutoff for dwarf galaxies. HALO7D can therefore provide new corroboration of the \citet{berg2012} results up to $z\sim1$, and for the other studies down to masses of $\sm \sim 8$. Using Spearman correlation tests, in the HALO7D low-mass sample, we find no $n_e$ dependence on SFR and only very weak correlations with stellar mass ($r=0.08$, $p=0.009$), sSFR ($r=-0.07$, $p=0.027$), and O32 ratio ($r=-0.1$, $p=0.010$). This is consistent with the studies collected in \citet{isobe2023} at this redshift range, which also find no notable dependence of $n_e$ on these properties in the combined sample. Additionally, we find no significant correlation with the SF surface density. 

At $1<z<3$, \citet{isobe2023} note a range in density of $100 < n_e < 250$ cm$^{-3}$ \citep[collected from][]{steidel2014, sanders2016, kaasinen2017, kashino2017, davies2021}, placing the HALO7D galaxies on the very low-density end at this redshift range ($n_e \sim 100$ cm$^{-3}$). However, the $z>1$ galaxies used in the \citet{isobe2023} composite are typically more massive ($\sm \sim 10$) than the HALO7D galaxies, so the comparison is not direct. 

\citet{kaasinen2017} suggested that the higher $n_e$ offset at $z>1$ compared to $z<1$ can be explained by higher SF in the detected samples at higher redshift. This is a plausible explanation for the offset seen in HALO7D. While it has been shown that HALO7D is complete down to the star-forming main sequence for $z<1$ dwarf galaxies \citep{pharo2022}, this is not necessarily the case for higher redshift dwarfs, which were not prioritized in candidate selection. This is reflected in the elevated median sSFR of the $z>1$ sample. 

For the $z<1$ dwarf sample where HALO7D probes a deeper SF range, we find that the dwarf galaxies have electron densities consistent with $z\sim0$ star forming galaxies. This echoes the findings for the gas-phase metallicity (see \S \ref{sec:strong}, SFR \citep{pharo2022}, and ionization properties \citep{pharo2023} for dwarf galaxies in HALO7D. This could result from several potential explanations. The first relates to depth of observation, as discussed above: surveys targeting $z>1$ are very limited in their ability to probe dwarf galaxies, often resulting in samples containing only substantial starbursts \citep[e.g., ][]{zeimann2015}. If SF activity influences metallicity measurements, even weakly, this could explain the divergence. Improved depth of rest-optical observations with JWST at this redshift may address this.

Second, evolution in stellar metallicity may influence ISM conditions through changes in the hardness of the incident ionizing spectrum. This has been modeled extensively for diagnostic ISM emission line ratios \citep[e.g., ][]{steidel2014, sanders2016, strom2017, jeong2020}. Analysis of the Ne3O2 line ratio in \citet{pharo2023} observed that median-stack HALO7D Ne3O2 ratios are consistent with higher stellar metallicities in photoionization models than median observations at $z \gtrsim 2$. The stellar metallicity of course does relate to the gas phase, but this may also lead to lower ionization states and temperatures in the ISM, changing observed line ratios and physical properties such as the density. A relation between stellar mass and stellar metallicity is observed in both the local \citep[e.g.,][]{kirby2013} and high-redshift universes \citep[e.g.,][]{cullen2019}, with changes in stellar metallicity linked to evolution in ionizing conditions such as the UV slope \citep{calabro2021}. \citet{cullen2019} find little stellar metallicity evolution from $z=5$ to $z=2.5$, but that this high-redshift sample is offset to low stellar Z compared to $z=0$. \citet{helton2022} observed massive galaxies at redshift comparable to HALO7D in the LEGA-C survey and found little evolution in nebular properties of massive galaxies from $z\sim1$ to $z\sim0$, potentially explicable by common stellar metallicity characteristics in this epoch. This interpretation suggests that the evolution in stellar metallicity, which then drives the evolution in observed ISM conditions, is primarily confined to the $1<z<2.5$ period of peak cosmic SFR.

Third, the characteristics of dwarf galaxies may change how they retain metals relative to more massive galaxies. \citet{concas2022} analyzed ionized outflows in SF galaxies at $z>1$, finding evidence that while outflows are clearly present in massive galaxies, they may occur more rarely and have less mass loading in dwarf galaxies, even at the epoch of peak SF. If this is so, dwarf galaxies may more efficiently retain metal-enriched gas, softening stellar radiation spectra and enhancing ISM cooling, which could more rapidly bring ISM conditions observed in dwarf galaxies in line with what is observed locally. Additionally, \citet{lin2020} find no significant dependence of sSFR on galaxy radius for dwarf SF galaxies at fixed stellar mass, suggesting that stellar winds may be weaker in smaller galaxies. This could result in more metal retention in dwarf galaxies, reflected in both the lack of strong correlation we find between Z or $RO2$ with $\Sigma_{SFR}$ and in the overall lower ionization conditions we find in $z<1$ dwarfs.

\newpage
\section{Summary and Conclusion} \label{sec:conc}

We have used deep optical Keck/DEIMOS spectroscopy from HALO7D and other surveys to calculate gas-phase metallicity measurements for 583 emission line galaxies with $0.35 < z < 0.85$, of which 388 are dwarf galaxies with $\sm < 9.5$. We construct MZRs for this sample using the O3H$\beta$, O3O2, and R23 strong-line ratios, finding good agreement with previous strong-line MZRs of low-mass galaxies at $z\sim0.7$, as well as with $z\sim0$ MZRs. We construct additional MZRs for 112 and 17 individual galaxies with \neiii\ and \auroral\ detections, respectively, as well as from measurements of composite spectra. These results yield MZRs more akin to $T_e$-derived metallicities at $1<z<3$, which are offset to lower metallicity at given mass than the $z=0$ MZR, in contradiction with the strong-line results. This could be a result of $T_e$-derived metallicities over-emphasizing higher temperature, lower Z regions, biasing the overall measurement in the composite spectra.

We estimate the intrinsic scatter in the metallicity in bins of stellar mass for each of the strong-line methods, finding a consistent trend of increased intrinsic scatter with lower stellar mass, supporting the proposition that dwarf galaxies are more susceptible to processes that may alter the chemical content of their nebular gas. We measure weak but significant correlations between metallicity offset from the best-fit MZR and the SFR, sSFR, and $\Sigma_{SFR}$ for the O3H$\beta$ and O3O2 strong-line methods, with lower metallicity offsets correlated with increased SF activity. We measure the FMR for the HALO7D sample, finding scatter is minimized with an SFR coefficient of $\alpha=0.21$, a weaker connection than is typically measured in the local Universe that corroborates other findings of weak SFR dependence at this redshift.

Finally, we use measurements of the \oii\ doublet to determine electron densities for HALO7D galaxies with $\sm < 10.5$ and $0.3<z<1.4$. We measure a median density of $n_e = 40$ cm$^{-3}$ for the entire low-mass sample, and find that these low densities, comparable to SF galaxies at $z=0$, hold for HALO7D low-mass galaxies with $z<1$. At $z>1$, we measure median densities of order 100 cm$^{-3}$ as well as a higher average sSFR. This echoes the findings of metallicity and ionization properties of HALO7D dwarfs, where the dwarf population resembles local galaxies for $z<1$. We discuss possible explanations for this, including increased stellar metallicity producing softer ionizing spectra and weaker, less mass-loaded winds in dwarf galaxies increasing metal retention in the interstellar gas.

\acknowledgments

We would like to thank the referee for their many detailed and helpful comments and suggestions. This research project was supported by NASA/ADAP grant No. 80NSSC20K0443. This research made use of Astropy, a community-developed core Python package for Astronomy \citep{astropy:2013, astropy:2018, astropy:2022}. We recognize and acknowledge the significant cultural role and reverence that the summit of Maunakea has always had within the indigenous Hawaiian community. We are most fortunate to have the opportunity to use observations conducted from this mountain.

\bibliography{refs}{}
\bibliographystyle{aasjournal}



\end{document}